\documentclass[12 pt]{article}
\usepackage{arxiv}
\usepackage{amsmath}
\numberwithin{equation}{section}
\usepackage{enumerate}
\usepackage{enumitem}
\usepackage[utf8]{inputenc} 
\usepackage[T1]{fontenc}    
\usepackage{hyperref}       
\usepackage{url}            
\usepackage{booktabs}       
\usepackage{amsfonts}       
\usepackage{nicefrac}       
\usepackage{microtype}      
\usepackage{amsbsy}   
\usepackage{natbib}   
\usepackage{cleveref}       
\usepackage{graphicx}   
\usepackage{longtable}
\usepackage{multirow}
\usepackage{indentfirst}   
\usepackage{here}
\usepackage{appendix}
\usepackage{fixmath}
\usepackage{float}
\usepackage{ulem}
\usepackage{bm} 
\usepackage{lscape} 
\usepackage{caption}
\usepackage{comment}

\usepackage{amssymb}

\usepackage[usenames,table,dvipsnames]{xcolor}
\usepackage{ae}
\usepackage{fancyvrb}
\usepackage{upquote}

\usepackage{textcomp}
\usepackage{listings}
\definecolor{myGray}{rgb}{0.75,0.75,0.75}
\usepackage{calc}

\usepackage[all]{hypcap}  
\usepackage{titletoc}

\colorlet{RED}{red}

\definecolor{referenceGreen}{rgb}{0.0, 0.75, 0.0}

\hypersetup{
    colorlinks=true,
    linkcolor=blue,
    filecolor=magenta,      
    urlcolor=blue,
    citecolor=referenceGreen,
    }

\makeatletter
\let\proglang=\textsf
\newcommand{\pkg}[1]{{\fontseries{b}\selectfont #1}}
\newcommand\code{\bgroup\@makeother\_\@makeother\~\@makeother\$\@codex}
\def\@codex#1{{\normalfont\ttfamily\hyphenchar\font=-1 #1}\egroup}

\DefineVerbatimEnvironment{Sinput}{Verbatim}{fontshape=sl}
\DefineVerbatimEnvironment{Soutput}{Verbatim}{}
\DefineVerbatimEnvironment{Scode}{Verbatim}{fontshape=sl}

\DefineVerbatimEnvironment{Code}{Verbatim}{}
\DefineVerbatimEnvironment{CodeInput}{Verbatim}{fontshape=sl}
\DefineVerbatimEnvironment{CodeOutput}{Verbatim}{}

\setkeys{Gin}{width=0.8\textwidth}

\definecolor{Red}{rgb}{0.5,0,0}
\definecolor{Blue}{rgb}{0,0,0.5}

\advance\textheight by \topskip
\makeatother

\begin{document}

\lstset{ 
  language=R,                     
  basicstyle=\footnotesize\ttfamily, 
  numbers=left,                   
  numberstyle=\footnotesize\color{Blue},  
  stepnumber=1,                   
  numbersep=5pt,                  
  backgroundcolor=\color{white},  
  showspaces=false,               
  showstringspaces=false,         
  showtabs=false,                 
  frame=single,                   
  rulecolor=\color{black},        
  tabsize=2,                      
  captionpos=b,                   
  breaklines=true,                
  breakatwhitespace=false,        
  keywordstyle=\color{RoyalBlue},      
  commentstyle=\color{YellowGreen},   
  stringstyle=\color{ForestGreen}      
}

\newlength\dlf
\newcommand\alignedbox[2]{
  &
  \begingroup
  \settowidth\dlf{$\displaystyle #1$}
  \addtolength\dlf{\fboxsep+\fboxrule}
  \hspace{-\dlf}
  \fcolorbox{white}{myGray}{$\displaystyle #1 #2$}
  \endgroup
}

\def\figsize{250px}
\def\verbatimfont{footnotesize}






\author{Silvaneo V. dos Santos Jr.\\Federal University\\ of Rio de Janeiro \And 
        Mariane Branco Alves\\Federal University\\ of Rio de Janeiro \And
        Hélio S. Migon\\Federal University\\ of Rio de Janeiro}
\title{\pkg{kDGLM}: a R package for Bayesian analysis of Generalized Dynamic Linear Models }
\shorttitle{\pkg{kDGLM}}
\rhead{\scshape \footnotesize \headeright}

\def\spacingset#1{\renewcommand{\baselinestretch}%
{#1}\small\normalsize} \spacingset{1}

\maketitle

\abstract{
This paper introduces \pkg{kDGLM}, an \proglang{R} package designed for Bayesian analysis of Generalized Dynamic Linear Models (GDLM), with a primary focus on both uni- and multivariate exponential families. Emphasizing sequential inference for time series data, the \pkg{kDGLM} package provides comprehensive support for fitting, smoothing, monitoring, and feed-forward interventions. The methodology employed by \pkg{kDGLM}, as proposed in \cite{ArtigokParametrico}, seamlessly integrates with well-established techniques from the literature, particularly those used in (Gaussian) Dynamic Models. These include discount strategies, autoregressive components, transfer functions, and more. Leveraging key properties of the Kalman filter and smoothing, \pkg{kDGLM} exhibits remarkable computational efficiency, enabling virtually instantaneous fitting times that scale linearly with the length of the time series. This characteristic makes it an exceptionally powerful tool for the analysis of extended time series. For example, when modeling monthly hospital admissions in Brazil due to gastroenteritis from 2010 to 2022, the fitting process took a mere 0.11s. Even in a spatial-time variant of the model (27 outcomes, 110 latent states, and 156 months, yielding 17,160 parameters), the fitting time was only 4.24s. Currently, the \pkg{kDGLM} package supports a range of distributions, including univariate Normal (unknown mean and observational variance), bivariate Normal (unknown means, observational variances, and correlation), Poisson, Gamma (known shape and unknown mean), and Multinomial (known number of trials and unknown event probabilities). Additionally, \pkg{kDGLM} allows the joint modeling of multiple time series, provided each series follows one of the supported distributions. Ongoing efforts aim to continuously expand the supported distributions.
}

\section{Introduction}\label{Sec:Intro}

The class of Dynamic Linear Models (DLMs) \citep{WestHarr-DLM, Petris-DLM} represents a flexible and computationally efficient approach to time series modeling. Utilizing the sequential nature of the fitting algorithm known as the Kalman Filter \citep{Kalman_filter_origins}, these models naturally accommodate monitoring tasks while allowing timely interventions by analysts. Although the original DLM framework is highly flexible, it has certain limitations. It is primarily applicable to problems where the observed outcome follows a Gaussian distribution, and the observational variance does not exhibit temporal dynamics. While some innovative solutions have been proposed to address the latter issue \citep[see][Chapter 10]{WestHarr-DLM}, these approaches do not support predictive structures for both the mean and observational variance.

As an extension of the original DLM framework, \cite{WestHarrMigon} introduced the Dynamic Generalized Linear Model (DGLM), a generalization that combines concepts from Generalized Linear Models (GLMs) \citep[GLM,][]{nelder1972generalized} and Dynamic Linear Models \citep{harrison1976bayesian}. The DGLM framework allows for sequential inference in cases where the outcome distribution belongs to the Exponential Family, but it still imposes the restriction that only one parameter, typically the mean, can have a predictive structure.

For non-Gaussian DGLMs, a general approach is to use some type of approximation, like Monte Carlo Markov chain (MCMC), which gives up the sequential aspect of those models and comes with the burden of high computational costs \citep{fruhwirth1994data, carter1994gibbs, gamerman1998markov, Shephard1997, Durbin2002}.

As for computational tools, the \pkg{dlm} \citep{DLM-pkg} offers a framework to fit Gaussian DLMs using Kalman Filter and Smoothing algorithms and, as such, preserving the sequential nature of DLMs and the low computational cost. The packages \pkg{KFAS} \citep{KFAS} and \pkg{FKF} \citep{FKF} offer the Kalman Filter and Smoother for Exponential Family State Space Models, but they doesn't directly compute the posterior distribution in the Bayesian sense, only providing tools for state estimation and forecasting in state space models. For a Bayesian analysis, the \pkg{bsts} \citep{bsts} package can be an option, as it is versatile in handling various types of time series data and incorporates Bayesian model averaging, which can be particularly powerful for predictive accuracy. Still, this packages relies on MCMC methods that, as mentioned, comes with high computational costs. Similarly, the \pkg{NGSSEML} \citep{gamerman2013non} package also provides tools for Bayesian inference in time series, even allowing for observational models outside the Exponential Family. For more general MCMC approachs, packages such as \pkg{rstan} \citep{RStan} and \pkg{rjags} \citep{rjags} serve as generic tools for Bayesian Models and, in particular, can be used for Dynamic Models. Lastly, the \pkg{rinla} \citep{RINLA, INLA} is a low cost alternative for time series modelling, although it does not have direct support for DLMs.

Returning to sequential approaches, a method similar to \cite{WestHarrMigon} was introduced in \cite{marotta2018} and formally proposed in \cite{ArtigokParametrico}. This approach leverages Bregman's projection theorem \citep{AmariGeomInfo} to derive a sequential algorithm for inference in $k$-parametric Exponential Families. While the posterior obtained through this method is approximate, it converges quickly to the exact posterior due to the properties of the Exponential Family.

Using this approach, one can fit models with multiple outcomes and predictive structures for all parameters of the observational model. Some specific cases covered by this method include univariate normal models with predictive structures for both the mean and observational variance, bivariate normal models with predictive structures for means, observational variances, and correlations, as well as multinomial models with an arbitrary number of categories. The sequential nature of this approach facilitates the incorporation of methodologies previously developed for Gaussian DLMs, such as discount strategies, autoregressive components, transfer functions, automated monitoring, and real-time interventions.

As a follow-up to the work presented in \cite{ArtigokParametrico}, we introduce the \proglang{R} package \pkg{kDGLM}. This package offers real-time inference capabilities for DGLMs, effectively translating the methods and concepts from \cite{WestHarr-DLM} and \cite{ArtigokParametrico} into practical applications.

Section \ref{Sec:Pkg} introduces the \pkg{kDGLM} package; this section highlights the mains of aspects package structure and offers insights into computational cost scaling. Section \ref{Sec:Model} provides a comprehensive overview of the general $k$-parametric Dynamic Generalized Linear Model (DGLM),  introducing the specific notation utilized throughout this paper and within the associated software package, additionally, it establishes key conventions and lays out the overarching framework used throughout the paper. In Sections \ref{Sec:Structure}, \ref{Sec:Outcome}, and \ref{Sec:Fitting}, we detail the key features of the \pkg{kDGLM} package, supplemented by extensive examples. It is recommended to read these sections sequentially, as each builds upon the concepts introduced in the preceding sections, mirroring the logical progression a new user would follow when learning to use \pkg{kDGLM}. However, for more experienced users, or upon subsequent readings, these sections can also serve as standalone reference points, particularly for those already familiar with the syntax.

Section \ref{Sec:Structure} focuses on defining the model's structure, linking latent states to linear predictors (see Section \ref{Sec:Model}), and discussing the temporal dynamics of the model. Section \ref{Sec:Outcome} delves into the specifics of the observational model and explores the handling of multiple outcomes. Finally, Section \ref{Sec:Fitting} presents the application of functions for fitting a model, utilizing the concepts and tools introduced in Sections \ref{Sec:Structure} and \ref{Sec:Outcome}. All sections are accompanied by practical, ready-to-use code examples.

Section \ref{Sec:casestudy} demonstrates a comprehensive application of the \pkg{kDGLM} package through a real-world case study: analyzing gastroenteritis incidents in Brazil from 2010 to 2022. This section methodically guides the reader through each step of utilizing the \pkg{kDGLM} package to address this issue. It begins with the implementation of a basic time series model and progressively advances to a more sophisticated Spatio-Temporal Model, including a Conditional Autoregressive (CAR) structure \citep{banerjee2014hierarchical, AlexCar}. The primary focus here is on the practical application of the \pkg{kDGLM} package and its diverse tools, highlighting its capabilities. While the case study is centered around an epidemiological issue, the section does not delve into the epidemiological aspects of gastroenteritis. Each step in the process is elucidated with ready-to-use code examples, enhancing the section's utility as a practical guide.

Finally, Section \ref{Sec:Conclusion} provides a summary of the paper's key insights. Additionally, it outlines prospective development pathways and future enhancements for the \pkg{kDGLM} package.

\section{The R package \pkg{kDGLM}}\label{Sec:Pkg}

\subsection{Dependencies}

The \pkg{kDGLM} package depends on \pkg{extraDistr} (>= 1.9.1), \pkg{cubature} (>=  2.1.0) and \pkg{Rfast} (>= 2.0.8). We also recommend the packages  \pkg{ggplot2} and \pkg{plotly}, which improve the built-in plot methods offered by the \pkg{kDGLM} package.

\subsection{Installation}

Currently, the \pkg{kDGLM} package is only available through the GitHub repository "silvaneojunior/kDGLM", but we are working to submit the first version to CRAN, which should happen soon. To install the most recent version of the \pkg{kDGLM} package, one can simply execute the following code line:

\begin{lstlisting}
devtools::install_github('silvaneojunior/kDGLM')
\end{lstlisting}

\subsection{Philosophy}

This package is designed for the construction, fitting, and analysis of $k$-parametric Dynamic Generalized Linear Models (DGLM), which can handle both multivariate outcomes and multiple outcomes simultaneously \citep[refer to][for more details on this distinction]{ArtigoMultivar}. We have crafted the syntax with a primary focus on facilitating its use in models where $k>1$, but we also provide support for uni-parametric models (such as the Poisson, Binomial, and Gamma with known shape). We have made efforts to keep the syntax as simple as possible without compromising its generality. Section \ref{Sec:Model} is dedicated to exploring the definition of $k$-parametric DGLMs, which, for simplicity, we will refer to as kDGLMs.

On an implementation level, this package is built upon two core principles: Online sequential inference and support for easy community contributions. The first principle is inherent from \cite{ArtigokParametrico}, whose methodology is particularly suited for real time inference and monitoring. While this package has support for a very wide range of applications, it is particularly suited for problems where data is observed continuously and analysts must make decisions on-the-fly; continuously learning and interacting with the system they monitor. An example of such problem could be the monitoring of deaths among patients of a hospital. The analyst could use this package to monitor the mortality rate, detecting any unexpected increase in mortality and/or measuring the effects of changes in the hospital policies. Therefore, our focus is on implementing tools adaptable to such dynamic contexts.

For the second principle, we chose to develop the \pkg{kDGLM} package in modules that interact with each order and can be easily expanded. Those modules are: (1) Structural module, (2) outcome module, (3) fitting module and (4) summary module. Figure \ref{fig:diagram} presents a visual scheme of this structure:

\begin{figure}[H]
    \centering
    \includegraphics{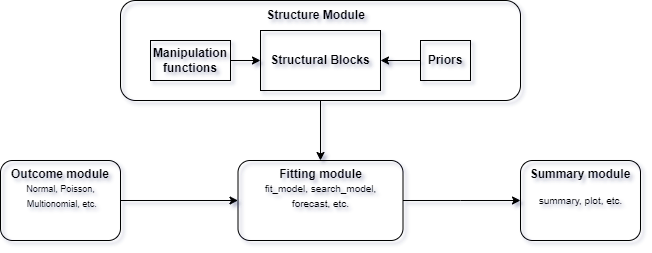}
    \caption{The structure of the \pkg{kDGLM} package and the relationship between modules. The direction of the arrows indicates the order the user must follow when using the package.}
    \label{fig:diagram}
\end{figure}

The \pkg{kDGLM} package is thoughtfully designed to promote seamless interoperability between its modules. This means that any function within one module can be used in conjunction with functions from any other module, creating a flexible and adaptable framework.

For example, suppose someone wishes to introduce a new type of structural block within the structure module. In such a case, this novel block should effortlessly integrate with any type of outcome and any fitting function. Similarly, if someone decides to introduce a new type of outcome in the outcome module, it should be compatible with any type of structural block and any fitting function. Additionally, functions responsible for summarizing fitted models must be capable of working with any combination of outcomes and structural components.

To guarantee this level of compatibility, each new function developed must adhere to the overarching structure of the specific module to which it belongs. This strategic approach is aimed at facilitating the integration of contributions from various sources over time, fostering collaboration, and streamlining the learning process for new developers who join the project. It underscores our commitment to creating an open and adaptable ecosystem for the benefit of the entire community.

\subsection{Computational cost}

By adopting the methodology proposed in \cite{ArtigokParametrico}, the \pkg{kDGLM} package inherits several key attributes reminiscent of the Kalman filter. These attributes, particularly relevant when considering the computational aspects of model fitting and forecasting, can be succinctly summarized as follows:

\begin{itemize}
    \item \textbf{Linear Time Scaling with Time Series Length or Forecast Range:} The computational time required for fitting and forecasting scales linearly with the length of the time series or the forecast range. This behavior arises from the sequential nature of the approach presented in \cite{ArtigokParametrico}. In essence, the computational cost of fitting a model with a time series of length $T$ is approximately $T$ times that of fitting the same model with just a single observation.

\item \textbf{Linear Time Scaling with the Number of Outcomes:} A similar linear scaling effect is observed concerning the number of outcomes in the model. Detailed information on this aspect can be found in \cite{ArtigoMultivar}.

\item \textbf{Quadratic Time Scaling with the Number of Parameters:} Lastly, the computational cost scales quadratically with the number of parameters incorporated into the model. This phenomenon is primarily attributed to the inversion of the covariance matrix required for the update algorithm. For a symmetric positive definite matrix, the computational cost of this inversion process scales quadratically with the dimension of the covariance matrix.
\end{itemize}

These computational scaling observations shed light on the performance characteristics of the \pkg{kDGLM} package, providing valuable insights into its computational efficiency and resource requirements. Moreover, they empower users to make informed estimates of the total fitting time based on smaller sub-sample.

\section{\texorpdfstring{$k-$}aParametric Dynamic Generalized Linear Models}\label{Sec:Model}

In this section, we assume the user's interest lies in analyzing a Time Series $\{\vec{Y}_t\}_{t=1}^T$, which adheres to the model described by \cite{ArtigokParametrico}:

\begin{equation}
\begin{aligned}
\vec{Y}_t|\vec{\eta}_t &\sim \mathcal{F}\left(\vec{\eta}_t\right),\\
g(\vec{\eta}_t) &=\vec{\lambda}_{t}=F_t'\vec{\theta}_t,\\
\vec{\theta}_t&=G_t\vec{\theta}_{t-1}+\vec{\omega}_t,\\
\vec{\omega}_t &\sim \mathcal{N}_n(\vec{h}_t,W_t),
\end{aligned}
\end{equation}

The model comprises:

\begin{itemize}
\item $\vec{Y}_t=(Y_{1,t},...,Y_{r,t})'$, the outcome, is an $r$-dimensional vector of observed variable.
\item $\vec{\theta}_t=(\theta_{1,t},...,\theta_{n,t})'$, representing the unknown parameters (latent states), is an $n$-dimensional vector, consistently dimensioned across observations.
\item $\vec{\lambda}_t=(\lambda_{1,t},...,\lambda_{k,t})'$, the linear predictors, is a $k$-dimensional vector indicating the linear transformation of the latent states. As per \cite{ArtigokParametrico}, $\vec{\lambda}_t$ is assumed to be (approximately) Normally distributed at all times and directly corresponds to the observational parameters $\vec{\eta}_t$, through a one-to-one correspondence $g$.
\item $\vec{\eta}_t=(\eta_{1,t},...,\eta_{l,t})'$, the observational parameters, is an $l$-dimensional vector defining the model's observational aspects. Typically, $l=k$, but this may not hold in some special cases, such as in the Multinomial model, where $k=l-1$.
\item $\mathcal{F}$, a distribution from the Exponential Family indexed by $\vec{\eta}_t$. Pre-determines the values $k$ and $l$, along with the link function $g$.
\item $g$, the link function, establishes a one-to-one correspondence between $\vec{\lambda}_t$ and $\vec{\eta}_t$.
\item $F_t$, the design matrix, is a user-defined, mostly known,  matrix of size $k \times n$.
\item $G_t$, the evolution matrix, is a user-defined, mostly known,  matrix of size $n \times n$.
\item $\vec{h}_t=(h_{1,t},...,h_{n,t})'$, the drift, is a known $n$-dimensional vector, typically set to $\vec{0}$ except for model interventions (refer to subsection \ref{intervention}).
\item $W_t$, a known covariance matrix of size $n \times n$, is specified by the user.
\end{itemize}

Per \cite{WestHarr-DLM}, we define $\mathcal{D}_t$ as the cumulative information after observing the first $t$ data points, with $\mathcal{D}_0$ denoting pre-observation knowledge of the process $\{Y_t\}^T_{t=1}$.

The specification of $W_t$ follows \cite{WestHarr-DLM}, section 6.3,  where $W_t=Var[G_t\theta_{t-1}|\mathcal{D}_{t-1}] \odot (1-D_t) \oslash D_t + H_t$. Here, $D_t$ (the discount matrix) is an $n \times n$ matrix with values between $0$ and $1$, $\odot$ represents the Hadamard product, and $\oslash$ signifies Hadamard division. $H_t$ is another known $n \times n$ matrix specified by the user. This formulation implies that if $D_t$ entries are all $1$, and $H_t$ entries are all $0$, the model equates to a Generalized Linear Model. 

A prototypical example within the general model framework is the Poisson model augmented with a dynamic level featuring linear growth and a single covariate $X$:

\begin{equation}
\begin{aligned}
Y_t|\eta_t &\sim \mathcal{P}\left(\eta_t\right),\\
ln(\eta_t) &=\lambda_{t}=\mu_t+\beta_t X_t,\\
\mu_t&=\mu_{t-1}+\beta_{t-1}+\omega_{\mu,t},\\
\nu_t&=\nu_{t-1}+\omega_{\nu,t},\\
\beta_t&=\beta_{t-1}+\omega_{\beta,t},\\
\omega_{\mu,t},\omega_{\nu,t},\omega_{\beta,t} &\sim \mathcal{N}_3(\vec{0},W_t),
\end{aligned}
\end{equation}

In this model, $\mathcal{F}$ denotes the Poisson distribution; the model dimensions are $r=k=l=1$; the state vector $\theta_t$ is $(\mu_t,\nu_t,\beta_t)'$ with dimension $n=3$; the link function $g$ is the natural logarithm; and the matrices $F_t$ and $G_t$ are defined as:

$$
F_t=\begin{bmatrix}
    1 \\
    0 \\
    X_t 
\end{bmatrix} \quad
G_t=\begin{bmatrix}
    1 & 1 & 0 \\
    0 & 1 & 0 \\
    0 & 0 & 1 
\end{bmatrix}
$$

Consider now a Normal model with unknown mean $\eta_{1,t}$ and unknown precision $\eta_{2,t}$:

\begin{equation}
\begin{aligned}
Y_t|\eta_t &\sim \mathcal{N}\left(\eta_{1,t},\eta_{2,t}^{-1}\right),\\
\eta_{1,t} &=\lambda_{1,t}=\mu_{1,t}+\beta_t X_t,\\
ln(\eta_{2,t}) &=\lambda_{2,t}=\mu_{2,t},\\
\mu_{1,t}&=\mu_{1,t-1}+\beta_{t-1}+\omega_{\mu_1,t},\\
\nu_t&=\nu_{t-1}+\omega_{\nu,t},\\
\beta_t&=\beta_{t-1}+\omega_{\beta,t},\\
\mu_{2,t}&=\phi_{t-1}\mu_{2,t-1}+\omega_{\mu_2,t},\\
\phi_{t}&=\phi_{t-1}+\omega_{\phi,t},\\
\omega_{\mu_1,t},\omega_{\nu,t},\omega_{\mu,t},\omega_{\beta,t},\omega_{\phi,t} &\sim \mathcal{N}_5(\vec{0},W_t),
\end{aligned}
\end{equation}

For this case, $\mathcal{F}$ represents the Normal distribution; the model dimensions are $r=1$ and $k=l=2$; the state vector $\theta_t$ is $(\mu_{1,t},\nu_t,\beta_t,\mu_{2,t},\phi_t)'$ with dimension $n=5$; the link function $g$ and matrices $F_t$, $G_t$ are:

$$
g\left(\begin{bmatrix}
    x_1 \\
    x_2
\end{bmatrix}\right)= \begin{bmatrix}
    x_1 \\
    \ln(x_2)
\end{bmatrix}\quad
F_t=\begin{bmatrix}
    1 & 0 \\
    0 & 0\\
    X_t & 0 \\
    0 & 1 \\
    0 & 0
\end{bmatrix} \quad
G_t=\begin{bmatrix}
    1 & 1 & 0 & 0 & 0\\
    0 & 1 & 0 & 0 & 0\\
    0 & 0 & 1 & 0 & 0\\
    0 & 0¨& 0 & \phi & 0\\
    0 & 0¨& 0 & 0 & 1
\end{bmatrix}
$$

This configuration introduces $l=2$ observational parameters, necessitating $k=2$ linear predictors. The first linear predictor pertains to the location parameter of the Normal distribution and includes a linear growth model and the covariate $X_t$. The second linear predictor, associated with the precision parameter, models log precision as an autoregressive (AR) process. We express this model in terms of an Extended Kalman Filter \citep[see][for details]{WestHarr-DLM, Kalman_filter_origins}. This formulation aligns with the concept of a traditional Stochastic Volatility model, as highlighted by \cite{ArtigokParametrico}.

Both the Normal and Poisson models illustrate univariate cases. However, the general model also accommodates multivariate outcomes, such as in the multinomial case. Consider a vector of counts $\vec{Y}_t=(Y_{1,t},Y_{2,t},Y_{3,t},Y_{4,t},Y_{5,t})'$, with $Y_{i,T} \in \mathbb{Z}$ and $N_t=\sum_{i=1}^{5}Y_{i,t}$. The model is:

\begin{equation}
\begin{aligned}
\vec{Y}_{5,t}|N_t,\vec{\eta}_{t} &\sim Multinomial\left(N_t,\vec{\eta}_{t}\right),\\
\ln\left(\frac{\eta_{i,t}}{\eta_{5,t}}\right) &=\lambda_{1,t}=\mu_{i,t},i=1,...,4\\
\mu_{i,t}&=\mu_{i,t-1}+\omega_{\mu_i,t},i=1,...,4\\
\omega_{\mu_1,t},\omega_{\mu_2,t},\omega_{\mu_3,t},\omega_{\mu_4,t} &\sim \mathcal{N}_4(\vec{0},W_t),
\end{aligned}
\end{equation}

In this multinomial model, $\mathcal{F}$ is the Multinomial distribution; the model dimensions are $r=5$, $l=5$ and $k=4$; the state vector $\theta_t$ is $(\mu_{1,t},\mu_{2,t},\mu_{3,t},\mu_{4,t})'$; $F_t$ and $G_t$ are identity matrices of size $4\times 4$; and the link function $g$ maps $\mathbb{R}^5$ in $\mathbb{R}^{4}$ as:

$$
g\left(\begin{bmatrix}
    x_1 \\
    x_2 \\
    x_3 \\
    x_4 \\
    x_5
\end{bmatrix}\right)= \begin{bmatrix}
    \ln\left(\frac{x_1}{x_5}\right) \\
    \ln\left(\frac{x_2}{x_5}\right) \\
    \ln\left(\frac{x_3}{x_5}\right) \\
    \ln\left(\frac{x_4}{x_5}\right)
\end{bmatrix}
$$

Note that in the Multinomial distribuition, $\eta_i\ge 0, \forall i$ and $\sum_{i=1}^{5} \eta_i=1$. Thus, only $k=l-1$ linear predictors are necessary to describe this model.

It's important to emphasize that while we have chosen to illustrate simple model structures, such as a random walk in the log odds for each outcome, neither the general model framework nor the \pkg{kDGLM} package restricts to these configurations. Analysts have the flexibility to tailor models to their specific contexts, including the incorporation of additional latent states to enhance outcome explanation.

Lastly, this general model framework can be extended to encompass multiple outcome models. For further details, see Appendix \ref{Appex:Outcomes}.

Given the complexity of manually specifying all model components, the \pkg{kDGLM} package includes a range of auxiliary functions to simplify this process. The subsequent section delves into these tools.

\section{Creating the model structure}\label{Sec:Structure}

In this section we will discuss the specification of the model structure. We will consider the structure of a model as all the elements that determine the relation between our linear predictor $\vec{\lambda}_t$ and our latent states $\vec{\theta}_t$ though time. Thus, the present section is dedicated  to the definition of the following, highlighted equations from a general dynamic generalized model:

\begin{equation}
\begin{aligned}
\vec{Y}_t|\vec{\eta}_t &\sim \mathcal{F}\left(\vec{\eta}_t\right),\\
g(\vec{\eta}_t) &= \fcolorbox{white}{myGray}{$\vec{\lambda}_{t}=F_t'\vec{\theta}_t \kern 0.65em ,$}\\
\alignedbox{\kern 0.2em\vec{\theta}_t }{=G_t\vec{\theta}_{t-1}+\vec{\omega}_t,}\\
\alignedbox{\vec{\omega}_t }{\sim \mathcal{N}_n(\vec{h}_t,W_t)\kern 0.35em .}
\end{aligned}
\end{equation}

Namely, we consider that the structure of a model consists of the matrices/vectors $F_t$, $G_t$, $\vec{h}_t$, $H_t$ and $D_t$.

Although we allow the user to manually define each entry of each of those matrices (which we \textbf{do not} recommend), we also offer tools to simplify this task. Currently, we offer support for the following base structures:

\begin{itemize}
    \item \code{polynomial_block}: Structural block for polynomial trends \citep[see][chapter 7]{WestHarr-DLM}. As special cases, this block has support for random walks and linear growth models.
    \item \code{harmonic_block}: Structural block for seasonal trends using harmonics \citep[see][chapter 8]{WestHarr-DLM}.
    \item \code{regression_block}: Structural block for (dynamic) regressions \citep[see][chapters 6 and 9]{WestHarr-DLM}.
    \item \code{AR_block}: Structural block for autoregressive components \citep[see][chapters 9 and 13]{WestHarr-DLM}.
    \item \code{noise_block}: Structural block for random effects \citep{ArtigoMultivar}.
\end{itemize}

For the sake of brevity, we will present only the details for the \code{polynomial_block}, since all other functions have very similar usage (the full description of each block can be found in the vignette, in the reference manual and in their respective help pages).

Along with the aforementioned functions, we also present some auxiliary functions and operations to help the user manipulate created structural blocks.

In Subsection \ref{Subsec:Poly_block}, we introduce the \code{polynomial_block} function. This section begins by examining a simplistic model, characterized by a single structural block and one linear predictor, with a completely known $F_t$ matrix. Subsection \ref{Subsec:mult_block} builds upon these concepts, exploring models that incorporate multiple structural blocks while maintaining a singular linear predictor. The focus shifts in Subsection \ref{Subsec:mult_lin_pred}, where we delve into the specification of multiple linear predictors within the same model. In Section \ref{Subsec:unknown_F}, the discussion turns to scenarios where $F_t$ includes one or more unknown components. Finally, Subsection \ref{Subsec:priors} provides a brief examination of functions used to define specialized priors.

\subsection{A structure for polynomial trend models}\label{Subsec:Poly_block}

\begin{lstlisting}
polynomial_block(..., order = 1, name = "Var.Poly",
                 D = 1, h = 0, H = 0,
                 a1 = 0, R1 = c(9, rep(1, order - 1)),
                 monitoring = c(TRUE, rep(FALSE, order - 1)))
\end{lstlisting}

Recall the notation introduced in Section \ref{Sec:Model} and revisited at the beginning of this section. The \code{polynomial_block} function will create a structural block based on \cite{WestHarr-DLM}, chapter 7. This involves the creation of a latent vector $\vec{\theta}_t=(\theta_{1,t},...,\theta_{n,t})'$, such that:

\begin{equation}
\begin{aligned}
\label{eq:def_pol}
\theta_{i,t} &= \theta_{i,t-1}+\theta_{i+1, t-1}+\omega_{i,t}, i=1,...,n-1\\
\theta_{n,t} &= \theta_{n,t-1}+\omega_{n,t},\\
\theta_1&\sim \mathcal{N}_k(a_1,R_1),\\
\omega_{1,t},...,\omega_{n,t}&\sim \mathcal{N}_n(\vec{h}_t,W_t),
\end{aligned}
\end{equation}

where $W_t=Var[\theta_t|\mathcal{D}_{t-1}]\odot (1-D_t) \oslash D_t+H_t$.

Let's dissect each component of this specification.

The \code{order} argument sets the polynomial block's order, correlating $n$ with the value passed.

The optional \code{name} argument aids in identifying each structural block in post-fitting analysis, such as plotting or result examination (see Section \ref{Sec:Fitting}).

The \code{D}, \code{h}, \code{H}, \code{a1}, and \code{R1} arguments correspond to $D_t$, $\vec{h}_t$, $H_t$, $\vec{a}_1$ and $R_1$, respectively.

\code{D} specifies the discount matrices over time. Its format varies: a scalar implies a constant discount factor; a vector of size $T$ (the length of the time series) means varying discount factors over time; a 
$n\times n$ matrix indicates that the same discount matrix is given by \code{D} and is the same for all times; a 3D-array of dimension $n\times n\times T$ indicates time-specific discount matrices. Any other shape for \code{D} is considered invalid.

\code{h} specifies the drift vector over time. If \code{h} is a scalar, it is understood that the drift is the same for all variables at all time. If \code{h} is a vector of size $T$, then it is understood that the drift is the same for all variables, but have different values for each time, such that each coordinate $t$ of \code{h} represents the drift for time $t$. If \code{h} is a $n \times T$ matrix, then we assume that the drift vector at time $t$ is given by \code{h[,t]}. Any other shape for \code{h} is considered invalid.

The argument \code{H} follows the same syntax as \code{D}, since the matrix $H_t$ has the same shape as $D_t$.

The argument \code{a1} and \code{R1} are used to define, respectively the mean and the covariance matrix for the prior for $\theta_1$. If \code{a1} is a scalar, it is understood that all latent states associated with this block have the same prior mean; if \code{a1} is a vector of size $n$, then it is understood that the prior mean $a_1$ is given by \code{a1}. If \code{R1} is a scalar, it is understood that the latent states have independent priors with the same variance (this does not imply that they will have independent posteriors); if \code{R1} is a vector of size $n$, it is understood that the latent states have independent priors and that the prior variance for the $\theta_{i,1}$ is given by \code{R1[i]}; if \code{R1} is a $n \times n$ matrix, it is understood that $R_1$ is given by \code{R1}.  Any other shape for \code{a1} or \code{R1} are considered invalid.

The arguments \code{D}, \code{h}, \code{H}, \code{a1}, and \code{R1} can accept character values, indicating that certain parameters are not fully defined. In such cases, the dimensions of these arguments are interpreted in the same manner as their numerical counterparts. For instance, if \code{D} is a single character, it implies a uniform, yet unspecified, discount factor across all variables and time points, with \code{D} serving as a placeholder label. Should \code{D} be a vector of length $T$ (the time series length), it suggests varying discount factors over time, with each character entry in the vector (e.g., \code{D[i]}) acting as a label for the discount factor at the respective time point. This logic extends to the other arguments and their various dimensional forms. It's crucial to recognize that if these arguments are specified as labels rather than explicit values, the corresponding model block is treated as "undefined," indicating the absence of a key hyperparameter. Consequently, a model with an undefined block cannot be fitted. Users must either employ the \code{set_block_value} function to replace labels with concrete values or use the \code{search_model} function to systematically evaluate models with different values for these labels. Section \ref{Subsec:sense_analysis} elaborates on the available tools for sensitivity analysis, including a detailed discussion on the use of \code{search_model}. Further information about both \code{set_block_value} and \code{search_model} is available in the reference manual or through the \code{help} function.

Notice that the user does not need to specify the matrix $G_t$, since it is implicitly determined by the equation \ref{eq:def_pol} and the order of the polynomial block. Each type of block will define it own matrix $G_t$, as such, the user does not need to worry about $G_t$, except in very specific circumstances, where an advanced user may need a type of model that is not yet implemented.


The argument \code{...} is used to specify the matrix $F_t$ (see details in Subsection \ref{Subsec:mult_lin_pred}). Specifically, the user must provide a list of named values which are arbitrary labels to each linear predictor $\lambda_{i,t}$ , $i=1,\ldots,k$, and its associated value represents the effect of the level $\theta_{1,t}$ (see Eq. (\ref{eq:def_pol})) in this predictor. 

For example, consider a polynomial block of order $2$, representing a linear growth. If the user passes an extra argument \code{lambda} (the naming is arbitrary) as $1$, then the matrix $F_t$ is created as:

$$
F_t=\begin{bmatrix}1\\0\end{bmatrix}
$$

Note that, as the polynomial block has order $2$, it has $2$ latent states, $\theta_{1,t}$ and $\theta_{2,t}$. While $\theta_{2,t}$ does not affect the linear predictor \code{lambda} directly, it serves as an auxiliary variable to induce a more complex dynamic for $\theta_{1,t}$. Indeed, by Equation \ref{eq:def_pol}, we have that a second order polynomial block have the following temporal evolution:

$$
\begin{aligned}
\theta_{1,t} &= \theta_{1,t-1}+\theta_{2, t-1}+\omega_{1,t}\\
\theta_{2,t} &= \theta_{2,t-1}+\omega_{2,t},\\
\omega_{1,t},\omega_{2,t}&\sim \mathcal{N}_2(\vec{h}_t,W_t).
\end{aligned}
$$

As such, $\theta_{2,t}$ represents a growth factor that is added in $\theta_{1,t}$ and smoothly changes overtime. Even more complex structures can be defined, either by a higher order polynomial block or by one of the several other types of block offered by the \pkg{kDGLM}.

The specification of values associated to each predictor label is further illustrated in the examples further exhibited in this section.  


Lastly, the argument \code{monitoring} shall be explained later, in Subsection \ref{intervention}, which discusses automated monitoring and interventions.


To exemplify the usage of this function, let us assume that we have a simple Normal model with known variance $\sigma^2$, in which $\eta$ is the mean parameter and the link function $g$ is such that $g(\eta)=\eta$. Let us also assume that the mean is constant over time and we have no explanatory variables, so that our model can be simply written as:

$$
\begin{aligned}
Y_t|\theta_t &\sim \mathcal{N}_1\left(\eta_t, \sigma^2\right),\\
\eta_t &=\fcolorbox{white}{myGray}{$\lambda_{t}=\theta_t,$}\\
\alignedbox{\theta_t}{=\theta_{t-1}=\theta.}
\end{aligned}
$$

In this case, we have $F_t=1$, $G_t=1$, $D_t=1$, $h_t=0$ and $H_t=0$, for all $t$. Assuming a prior distribution $\mathcal{N}(0,9)$ for $\theta$, we can create the highlighted structure using the following code:

\begin{lstlisting}
mean_block <- polynomial_block(eta = 1, order = 1, name = "Mean")
\end{lstlisting}

Setting \code{eta=1}, we specify that there is a linear predictor named \textit{eta}, and that $eta = 1 \times \theta$. Setting \code{order = 1}, we specify that $\theta_t$ is a scalar and that $G_t=1$. We can omit the values of \code{a1} , \code{R1}, \code{D}, \code{h} and \code{H}, since the default values reflect the specified model. We could also omit the argument \code{order}, since the default is $1$, but we chose to explicit define it so as to emphasize its usage. The argument \code{name} specifies a label for the created block; in this case, we chose to call it "Mean", to help identify its role in our model. 

Suppose now that we have an explanatory variable $X$ that we would like to introduce in our model to help explain the behavior of $\eta_t$. We could similarly define such structure by creating an additional block such as:

\begin{lstlisting}
polynomial_block(eta = X, name = "Var X")
\end{lstlisting}

By setting \code{eta=X}, we specify that there is a linear predictor called \textit{eta}, and that $eta = X \times \theta$. If $X=(X_1,...,X_T)'$ is a vector, then we would have $F_t=X_t$, for each $t$, such that $eta_t = X_t \times \theta$.

It should be noted that \pkg{kDGLM} has a specific structural block designed for regressions, \code{regression_block}, but we also allow any structural block to be used for a regression, by just setting the value assigned to the predictor equal to the regressor vector $X_t, t=1, \ldots, X_T$. 


The user can specify complex temporal dynamics for the effects of any co-variate. For instance, it could be assumed that a regressor has a seasonal effect on a linear predictor. This this could be accommodated by the insertion of the values of the regressor associated to a seasonal block. The use of seasonal blocks is illustrated in Section \ref{Sec:casestudy}.

So far, we have only discussed the creation of static latent effects, but the inclusion of stochastic temporal dynamics is very straightforward. One must simply specify the values of \code{H} to be greater than $0$ and/or the values of \code{D} to be lesser than $1$:

\begin{lstlisting}
mean_block <- polynomial_block(eta = 1, order = 1, name = "Mean", D = 0.95)
\end{lstlisting}

Notice that a dynamic regression model could be obtained by assigning \code{eta=X} in the previous code line. Bellow we present a plot of two simple trend models fitted to the same data: one with a static mean and another using a dynamic mean. 

In the following example we use the functions \code{Normal}, \code{fit_model} and the \code{plot} method for the \code{fitted_dlm} class. We advise the reader to initially concentrate solely on the application of the \code{polynomial_block}. The functionalities and detailed usage of the other functions, \code{Normal}, \code{fit_model}, and \code{plot}, will be explored in later sections, specifically in Sections \ref{Sec:Outcome} and \ref{Sec:Fitting}. The inclusion of these functions in the current example is primarily to offer a comprehensive and operational code sample.

\begin{lstlisting}
# Creating artifitial data
T <- 200
mu <- rnorm(T, 0, 0.5)
data <- rnorm(T, cumsum(mu))

# Creating structure
level1 <- polynomial_block(mu1 = 1, D = 1, name = "Static mean")
level2 <- polynomial_block(mu2 = 1, D = 0.95, name = "Dynamic mean")

fitted.model <- fit_model(level1, level2,
  Static.mean = Normal(mu = "mu1", V = 1, data = data),
  Dynamic.mean = Normal(mu = "mu2", V = 1, data = data))

plot(fitted.model, lag = -1)
\end{lstlisting}

\begin{figure}[H]
    \centering
    \includegraphics[width=\figsize]{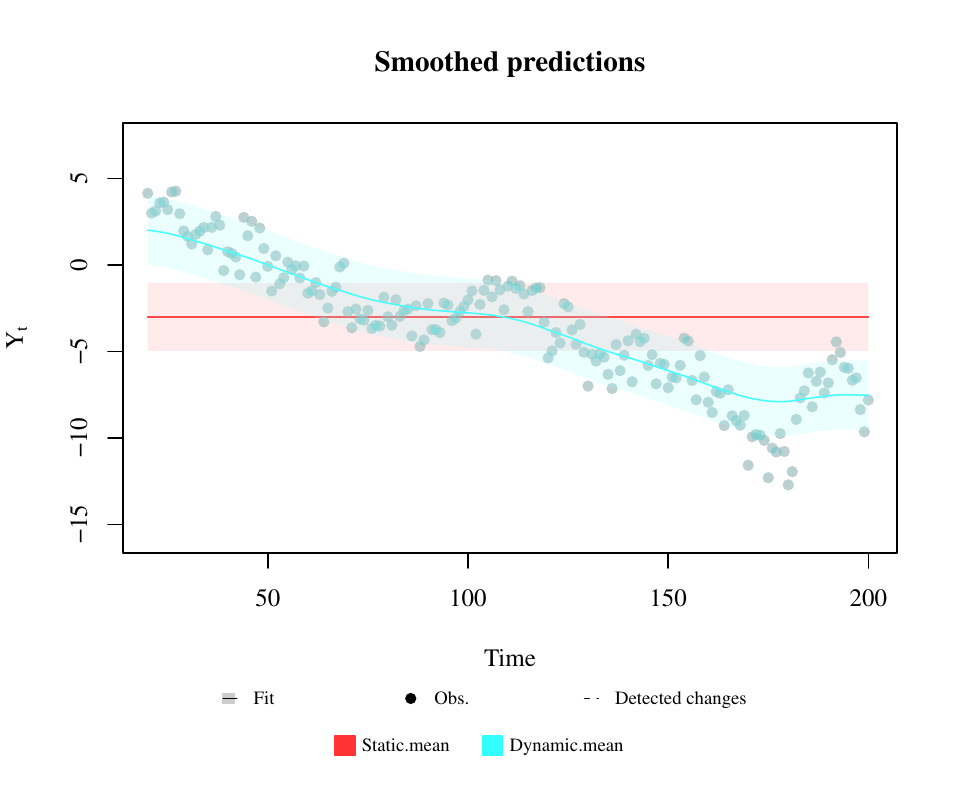}
    \label{fig:plot1}
\end{figure}

For an extensive presentation and thorough discussion of the theoretical aspects underlying the structure highlighted in this section, interested readers are encouraged to consult \citet[][ Chapters 6, 7, and 9]{WestHarr-DLM}. Additionally, we strongly recommend that all users refer to the associated documentation for more detailed information. This can be accessed by using the \code{help(polynomial_block)} function or consulting the reference manual. 

\subsection{Handling multiple structural blocks}\label{Subsec:mult_block}

In the previous subsection, we discussed how to define the structure of a model using the function \code{polynomial_block}. In a similar fashion, one could use the \code{regression_block}, \code{harmonic_block}, \code{AR_block} and \code{noise_block} functions to create different types of structures. Details on the use of these functions can be seen in the vignette, reference manual and their respective documentation. Each of these functions results in a single structural block. Generally, the user will want to mix multiple types of structures, each one being responsible to explain part of the outcome $Y_t$. For this task, we introduce an operator designed to combine structural blocks by superposition principle \citep[see][Sec. 6.2]{WestHarr-DLM}, as follows.

Consider the scenario where one wishes to superimpose $p$ structural blocks; for instance: trend, seasonal and regression components ($p=3$). A general overlaid structure is given by the following specifications:
$$
\begin{aligned}
\vec{\theta}_t&=\begin{bmatrix}\vec{\theta}_t^1\\ \vdots\\ \vec{\theta}_t^n\end{bmatrix}, &
F_t&=\begin{bmatrix}F_t^1 \\ 
\vdots \\ 
F_t^p\end{bmatrix},\\
G_t&=diag\{G_t^{1},...,G_t^{p}\},&
W_t&=diag\{W_t^{1},...,W_t^{p}\},
\end{aligned}
$$

where $diag\{M^1,...,M^{p}\}$ represents a block diagonal matrix such that its diagonal is composed of  $M^1,...,M^{p}$; $\theta_t$ is the vector obtained by the concatenation of the vectors $\vec{\theta}_t^1,..., \vec{\theta}_t^p$ corresponding to each structural block; and $F_t$ is obtained as follows: if a single linear predictor is considered in the model, $F_t$ is a line vector concatenating $F_t^1,..., F_t^p $. For the case of several predictors  ($k>1$, which will be seen in the next section), the design matrix associated to  structural block $i$,  $F_t^i$, has dimension $n_i \times k $ and $F_t$ is a $n \times k$ matrix,  obtained by the row-wise concatenation of the matrices $F_t^1,..., F_t^p$,  where $n=\sum_{i=1}^p n_i$.

In this scenario, to facilitate the specification of such model, we could create one structural block for each $\vec{\theta}_t^i$, $F_t^{i}$, $G_t^{i}$ and $W_t^{i}$, $i=1,...p$, and then "combine" all blocks together.  The \pkg{kDGLM} package allows this operation through the function \code{block_superpos} or, equivalently, through the \code{+} operator:

\begin{lstlisting}
block_1 <- ...
.
.
.
block_n <- ...

complete_structure <- block_superpos(block_1, ..., block_n)
# or
complete_structure <- block_1 + ... + block_n
\end{lstlisting}

For a very high number $p$ of structural blocks, the use of \code{block_superpos} is slightly faster. To demonstrate the usage of the \code{+}  operator, suppose we would like to create a model using four of the structures presented previously (a polynomial trend, a dynamic regression, a harmonic trend and an AR model). We could do so with the following code:

\begin{lstlisting}
poly_subblock <- polynomial_block(eta = 1,  name = "Poly",  D = 0.95)

regr_subblock <- regression_block(eta = X,  name = "Regr",  D = 0.95)

harm_subblock <- harmonic_block(eta = 1,  period = 12,  name = "Harm")

AR_subblock <- AR_block(eta = 1,  order = 1,  noise.var = 0.1,  name = "AR")

complete_block <- poly_subblock + regr_subblock + harm_subblock + AR_subblock
\end{lstlisting}

In the multiple regression context, that is, if more than one regressor should be included in a predictor, the user must specify different regression sub blocks, one for each regressor, and apply the superposition principle to these blocks. Thus, in the previous code lines, \code{X} is a vector with $T$ observations of a regressor $X_t$, already defined in an \proglang{R} object in the current enviroment and cannot be a matrix of covariates.  Ideally, the user should also provide each block with a name to help identify them after the model is fitted, but, if the user does not provide a name, the block will have the default name for that type of block. If different blocks have the same name, an index will be automatically added to the variables with conflicting labels based on the order that the blocks were combined. Note that the automatic naming might make the analysis of the fitted model confusing, specially when dealing with a large number of latent states. With that in mind, we \textbf{strongly} recommend the users to specify an intuitive name for each structural block.

When integrating multiple blocks within a model, it's crucial to understand how their associated design matrices, denoted as $F_t^{i}$ for each block, are combined. These matrices are concatenated vertically, one below the other. Consequently, since the predictor vector $\vec{\lambda}_t$ is calculated as $F_{t}'\vec{\theta}_t$, the influence of each block on $\vec{\lambda}_t$ is cumulative. In our previous code example, we introduced a linear predictor named \code{eta}. In this context, the operations performed in lines \code{1}, \code{5}, and \code{7} (corresponding to \code{polynomial_block}, \code{regression_block}, and \code{AR_block}, respectively), are represented as $eta_t=1 \times \theta_{1,t}^{i},i=1,3,4$; while in line \code{3} (corresponding to \code{regression_block}), the operation is $eta_t=X_t \times \theta_{1,t}^{2}$. It's important to note that each block initially defines \code{eta} independently. However, when these blocks are combined, their respective equations are merged. As a result, the complete structure in line \code{9} can be expressed as:

$$
eta_t= 1 \times \theta_{1,t}^{1} + X_t \times \theta_{1,t}^{2} + 1 \times \theta_{1,t}^{3} + 1 \times \theta_{1,t}^{4}
$$

This expression illustrates how the contributions from each individual block are aggregated to form the final model. This methodology allows for the flexible construction of complex models by combining simpler components, each contributing to explain a particular facet of the process $\{Y_t\}_{t=1}^T$.

\subsection{Handling several linear predictors}\label{Subsec:mult_lin_pred}

As the user may have noticed, more then one argument can be passed in the  \code{...} argument. Indeed, if the user does so, several linear predictors will be created in the same block (one for each unique name), all of which are affected by the associated latent state. For instance, take the following code:

\begin{lstlisting}
polynomial_block(lambda1=1,lambda2=1,lambda3=1) # Common factor
\end{lstlisting}

The code above creates $3$ linear predictors $\lambda_{1,t}$,$\lambda_{2,t}$ and $\lambda_{3,t}$ and a design matrix $F_t=(1,1,1)'$, such that:

$$
\begin{aligned}
\lambda_{1,t}&=1 \times \theta_{t}\\
\lambda_{2,t}&=1 \times \theta_{t}\\
\lambda_{3,t}&=1 \times \theta_{t}
\end{aligned}
$$

Note that the latent state $\theta_{t}$ is the same for all linear predictors $\lambda_{i,t}$, i.e., $\theta_{t}$ is a shared effect among those linear predictors which could be used to induce association among predictors. The specification of independent effects  to each linear predictor can be done by using different blocks to each latent state:

\begin{lstlisting}
polynomial_block(lambda1=1,order=1)+ # theta_1
polynomial_block(lambda2=1,order=1)+ # theta_2
polynomial_block(lambda3=1,order=1) # theta_3
\end{lstlisting}

When the name of a linear predictor is missing from a particular block, i.e., the name of a linear predictor was passed as an argument in one block, but is absent in another, it is understood that particular block has no effect on the linear predictor that is absent, such that the previous code would be equivalent to:

\begin{lstlisting}
# Longer version of the previous code for the sake of clarity.
# In general, when a block does not affect a particular linear predictor, that linear predictor should be ommited when creating the block.
polynomial_block(lambda1=1,lambda2=0,lambda3=0,order=1)+ # theta_1
polynomial_block(lambda1=0,lambda2=1,lambda3=0,order=1)+ # theta_2
polynomial_block(lambda1=0,lambda2=0,lambda3=1,order=1) # theta_3
\end{lstlisting}

As discussed in the end of subsection \ref{Subsec:mult_block}, the effect of each block over the linear predictors will be added to each other. As such both codes will create 3 linear predictors, such that:

$$
\begin{aligned}
\lambda_{1,t}&=1 \times \theta_{1,t} + 0 \times \theta_{2,t} + 0 \times \theta_{3,t}=\theta_{1,t}\\
\lambda_{2,t}&=0 \times \theta_{1,t} + 1 \times \theta_{2,t} + 0 \times \theta_{3,t}=\theta_{2,t}\\
\lambda_{3,t}&=0 \times \theta_{1,t} + 0 \times \theta_{2,t} + 1 \times \theta_{3,t}=\theta_{3,t}
\end{aligned}
$$

Remind the sintax presented in the first illustration of the current section, which guides the creation of common factors among predictors. One can use multiple blocks in the same structure to define linear predictors that share some (but not all) of their components:

\begin{lstlisting}
polynomial_block(lambda1=1,order=1)+ # theta_1
polynomial_block(lambda2=1,order=1)+ # theta_2
polynomial_block(lambda3=1,order=1)+ # theta_3
polynomial_block(lambda1=1,lambda2=1,lambda3=1,order=1) # theta_4: Common factor
\end{lstlisting}
representing the following structure:

$$
\begin{aligned}
\lambda_{1,t}&=\theta_{1,t}+\theta_{4,t}\\
\lambda_{2,t}&=\theta_{2,t}+\theta_{4,t}\\
\lambda_{3,t}&=\theta_{3,t}+\theta_{4,t}\\
\end{aligned}
$$

The examples above all have very basic structures, so as to not overwhelm the reader with overly intricate models. Still, the \pkg{kDGLM} package is not limited to in any way by the inclusion of multiple linear predictors, such that any structure one may use with a single predictor can also be used with multiple linear predictors. For example, we could have a model with $3$ linear predictors, each one having a mixture of shared components and exclusive components:

\begin{lstlisting}
#### Global level with linear growth ####
polynomial_block(lambda1=1,lambda2=1,lambda3=1,D=0.95,order=2)+
#### Local variables for lambda1 ####
polynomial_block(lambda1=1,order=1)+
regression_block(lambda1=X1,max.lag=3)+
harmonic_block(lambda1=1,period=12,D=0.98)+
#### Local variables for lambda2 ####
polynomial_block(lambda2=1,order=1)+
AR_block(lambda2=1,pulse=X2,order=1,noise.disc=1)+
harmonic_block(lambda2=1,period=12,D=0.98,order=2)+
#### Local variables for lambda3 ####
polynomial_block(lambda3=1,order=1)+
AR_block(lambda3=1,order=2,noise.disc=0.9)+
regression_block(lambda3=X3,D=0.95)
\end{lstlisting}

Now we focus on the replication of structural blocks, for which we apply \code{block_mult} function and the associated operator \code{*}. This function allows the user to create multiple blocks with identical structure, but each one being associated with a different linear predictor. The usage of this function is as simple as:

\begin{lstlisting}
base.block <- polynomial_block(eta = 1,  name = "Poly",  D = 0.95,order=1)

# final.block <- block_mult(base.block, 4)
# or
# final.block <- base.block * 4
# or
final.block <- 4 * base.block
\end{lstlisting}

When replicating blocks, it is understood that each copy of the base block is independent of each other (i.e., they have their own latent states) and each block is associated with a different set of linear predictors. The name of the linear predictors associated with each block are taken to be the original names with an index:

\begin{lstlisting}
final.block$pred.names
\end{lstlisting}
\begin{\verbatimfont}
\begin{verbatim}
[1] "eta.1" "eta.2" "eta.3" "eta.4"
 \end{verbatim}
 \end{\verbatimfont}

Naturally, the user might want to rename the linear predictors to a more intuitive label. For such task, we provide the \code{rename_block} function:

\begin{lstlisting}
final.block <- block_rename(final.block,c("Matthew","Mark","Luke","John"))
final.block$pred.names
\end{lstlisting}
\begin{\verbatimfont} \begin{verbatim}
[1] "Matthew" "Mark"    "Luke"    "John"   
 \end{verbatim}\end{\verbatimfont}

\subsection{Handling unknown components in the planning matrix $F_t$}\label{Subsec:unknown_F}

In some situations the user may want to fit a model such that:

$$
\begin{aligned}
\lambda_{t}=F_t'\theta_t=\cdots+\phi_t\theta_t +\cdots,
\end{aligned}
$$
in other words, it may be the case that the planning matrix $F_t$ contains one or more unknown components. This idea may be foreign when working with only one linear predictor, but if our observational model has several predictors it could make sense to have shared effects among predictors. Besides, this construction is also natural when modeling multiple time series simultaneously, such as when dealing with correlated outcomes or when working with a compound regression. All those cases will be explored in the Advanced Examples section of the vignette. For now, we will focus on \textbf{how} to specify such structures, whatever their use may be.

For simplicity, let us assume that we want to create a linear predictor $\lambda_t$ such that $\lambda_{t}=\phi_t\theta_t$. Then the first step would be to create a linear predictor associated with $\phi_t$ (which we will call \code{phi}, although the user may call it whatever it pleases the user):

\begin{lstlisting}
phi_block=polynomial_block(phi=1,order=1)
\end{lstlisting}

Notice that we are creating a linear predictor $\phi_t$ and a latent state $\tilde{\theta}_t$ such that $\phi_t=1\times \tilde{\theta}_t$. Also, it is important to note that the structure for $\phi_t$ could be any other structural block (harmonic, regression, auto regression, etc.).

Now we can create a structural block for $\theta_t$:

\begin{lstlisting}
theta_block=polynomial_block(lambda='phi',order=1)
\end{lstlisting}

The code above creates a linear predictor $\lambda_t$ and a latent state $\theta_t$ such that $\lambda_t=\phi_t \times \theta_t$. Notice that the \code{...} argument of any structural block is used to specify the planning matrix $F_t$, specifically, the user must provide a list of named values, where each name indicates a linear predictor $\lambda_t$ and its associated value represent the effect of $\theta_{t}$ in this predictor. When the user pass a string in \code{...}, it is implicitly that the component of $F_t$ associated with $\theta_t$ is unknown and modelled by the linear predictor labelled as the passed string.

Lastly, as one could guess, it is possible to establish a chain of components in $F_t$ in order to create an even more complex structure. For instance, take the following code:

\begin{lstlisting}
polynomial_block(eta1=1,order=1)+
polynomial_block(eta2='eta1',order=1)+
polynomial_block(eta3='eta2',order=1)
\end{lstlisting}

In the first line we create a linear predictor $\eta_{1,t}$ such that $\eta_{1,t}=1 \times \theta_{1,t}$. In the second line we create another linear predictor $\eta_{2,t}$ such that $\eta_{2,t}=\eta_{1,t} \times \theta_{2,t}=\theta_{1,t} \times \theta_{2,t}$. Then we create a linear predictor $\eta_{3,t}$ such that $\eta_{3,t}=\eta_{2,t} \times \theta_{3,t}=\theta_{1,t} \times \theta_{2,t} \times \theta_{3,t}$.

To fit models with non-linear components in the $F_t$ and/or $G_t$ matrices, we use the Extended Kalman Filter \citep[see][for details]{WestHarr-DLM,Kalman_filter_origins}.

\subsection{Special priors}\label{Subsec:priors}

The user may want to specify some special priors that impose a certain structure for the data. For instance, the user may believe that a certain set of latent state sum to $0$ or that there is a spacial structure to them. This is specially relevant when modelling multiple time series, for instance, lets say that we have $r$ series $Y_{i,t}$, $i=1,...r$, such that:

$$
\begin{aligned}    
Y_{i,t}|\eta_{i,t} &\sim Poisson(\eta_{i,t})\\
\ln(\eta_{i,t})&=\lambda_{it}=\mu_t+\alpha_{i,t},\\
\sum_{i=1}^{r} \alpha_{i,t}&=0, \forall t.
\end{aligned}
$$

Similarly, one could want to specify a CAR prior \citep{AlexCar, banerjee2014hierarchical} for the variables $\alpha_1,...\alpha_r$, if the user believes there is spacial autocorrelation.

For those scenarios, the \pkg{kDGLM} package offers some functions to modify the prior of structural blocks, such as the \code{zero_sum_prior} and the \code{CAR_prior}. Their general usage is very similar and can be simply used as:

\begin{lstlisting}
structure=polynomial_block(mu=1,D=0.95) |> block_mult(5) |> zero_sum_prior()
\end{lstlisting}

\section{Creating the model outcome}\label{Sec:Outcome}

We have presented the tools for creating the structure of a DGLM model, specifically, we have shown how to define the relationship between the latent vector $\vec{\theta}_t$ and the linear predictors $\vec{\lambda}_t$, along with the temporal dynamic of $\vec{\theta}_t$. Now we proceed to define the observational model for $\vec{Y}_t$ and the relationship between $\vec{\lambda}_t$ and $\vec{\eta}_t$, i.e., the highlighted part of the following equations:

\begin{align*}
\alignedbox{\kern 0.1em \vec{Y}_t|\vec{\eta}_t }{\sim \mathcal{F}\left(\vec{\eta}_t\right),}\\
 \alignedbox{g(\vec{\eta}_t) }{= \vec{\lambda}_{t}}=F_t'\vec{\theta}_t,\\
\vec{\theta}_t &=G_t\vec{\theta}_{t-1}+\vec{\omega}_t,\\
\vec{\omega}_t &\sim \mathcal{N}_n(\vec{h}_t,W_t),
\end{align*}

In each subsection, we will assume that the linear predictors are already defined, along with all the structure that comes along with them (i.e., we will take for granted the part of the model that is not highlighted), moreover, we also assume that the user has created the necessary amount of linear predictors for each type of outcome and that those linear predictors were named as $\lambda_1$,...,$\lambda_k$.

Currently, we offer support for the following observational distributions:

\begin{itemize}
    \item Normal distribution with unknown mean and unknown variance (with dynamic predictive structure for both parameters). As a particular case, we also have support for Normal distribution with known variance.
    \item Bivariate Normal distribution with unknown means, unknown variances and unknown correlation (with dynamic predictive structure for all parameters). As a particular case, we also have support for Multivariate Normal distribution with known covariance matrix.
    \item Poisson distribution with unknown rate parameter with dynamic predictive structure.
    \item Multinomial distribution with an known number of trials, arbitrary number of categories, but unknown event probabilities with dynamic predictive structure for the probability of each category. As particular cases, we support the Binomial and Bernoulli distributions.
    \item Gamma distribution with known shape parameter, but unknown mean with dynamic predictive structure.
\end{itemize}

We are currently working to include several distributions. In particular, the following distributions shall be supported very soon: Dirichlet; Geometric; Negative Binomial; Rayleigh; Pareto; Asymmetric Laplace with known mean.

For the sake of brevity, we will present only the Normal case, since the other cases have very similar usage (see the vignette, the reference manual and/or their
respective help pages for detailed information on each distribution).

\subsection{Creating Normal outcomes}

In some sense, we can think of this as the most basic case, at least in a theoretical point of view, since the Kalman Filter was first developed for this specific scenario \citep{Kalman_filter_origins}. Indeed, if we have a \textbf{static} observational variance/covariance matrix (even if unknown), we fall within the DLM class, which has an exact analytical solution for the posterior of the latent states. With some adaptations, one can also have some degree of temporal dynamic for the variance/covariance matrix \citep[see][section 10.8]{ameen1985discount, WestHarr-DLM}. Yet, the \pkg{kDGLM} package goes a step further, offering the possibility for predictive structure for both the mean and the observational variance/covariance matrix, allowing the inclusion of dynamic regressions, seasonal trends, autoregressive components, etc., for \textbf{both} parameters.

We will present this case in two contexts: the first, which is a simple implementation of the Kalman Filter and Smoother, deals with data coming from an Normal distribution (possibly multivariate) with unknown mean and known variance/covariance matrix; the second deals with data coming from a univariate Normal distribution with unknown mean and unknown variance.

Also, at the end of the second subsection, we present an extension to the bivariated Normal distribution with unknown mean and unknown covariance matrix. A study is being conducted to expand this approach to the $k$-variated case, for any arbitrary $k$.

\subsubsection{Normal outcome with known variance}

Suppose that we have a sequence of $k$-dimensional vectors $\vec{Y}_t$, such that $\vec{Y}_t=(Y_{1t},...,Y_{kt})'$. We assume that:

$$
\begin{aligned}
\vec{Y}_t|\mu_t,V &\sim \mathcal{N}_k\left(\vec{\mu}_t,V\right),\\
\mu_{it}&=\lambda_{it}, i=1,...,k,\\
\end{aligned}
$$
where $\vec{\mu}_t=(\mu_{1,t},...,\mu_{k,t})'$ and $V$ is a known symmetric, definite positive $k\times k$ matrix. Also, for this model, we assume that the link function $g$ is the identity function.

To create the outcome for this model, we can make use of the \code{Normal} function:

\begin{lstlisting}
Normal(mu, V = NA, Tau = NA, Sd = NA, data)
\end{lstlisting}

Intuitively, the \code{mu} argument must be a character vector of size $k$ containing the names of the linear predictors associated with each $\mu_{i.}$. The user must also specify one (and only one) of \code{V}, \code{Tau} or \code{Sd}. If the user provides \code{V}, $V$ is assumed to be that value; if the user provides \code{Tau}, $V$ is assumed to be the inverse of the given matrix (i.e., \code{Tau} is the precision matrix); if the user provides \code{Sd}, $V$ is assumed to be such that the standard deviation of the observations is equal to the main diagonal of \code{Sd} and the correlation between observations is assumed the be equal to the off-diagonal elements of \code{Sd}.

The \code{data} argument must be a $T \times k$ matrix containing the values of $\vec{Y}_t$ for each observation. Notice that each line $t$ must have the values of all categories in time $t$ and each column $i$ must represent the values of a category $i$ through time. If a value of the argument \code{data} is not available (\code{NA}) for a specific time, it is assumed that there was no observation at that time, thus the update step of the filtering algorithm will be skipped at that time. Note that the evolution step will still be performed, such that the predictive distribution for the missing data and the updated distribution for the latent states at that time will still be provided.

Next, we present a brief example for the usage of \code{Normal} function for a univariate outcome (the multivariate case works similarly). We use some functions described in the previous sections, as well as some functions that will be presented later on. For now, let us focus on the usage of the \code{Normal} function.

\begin{lstlisting}
level <- polynomial_block(mu = 1, D = 0.95, order = 2)
season <- harmonic_block(mu = 1, period = 12, D = 0.975)

outcome <- Normal(mu = "mu", V = 6e-3,
                  data = c(log(AirPassengers)))
fitted.model <- fit_model(level,season, outcome)
plot(fitted.model)
\end{lstlisting}
\begin{figure}[H]
    \centering
    \includegraphics[width=\figsize]{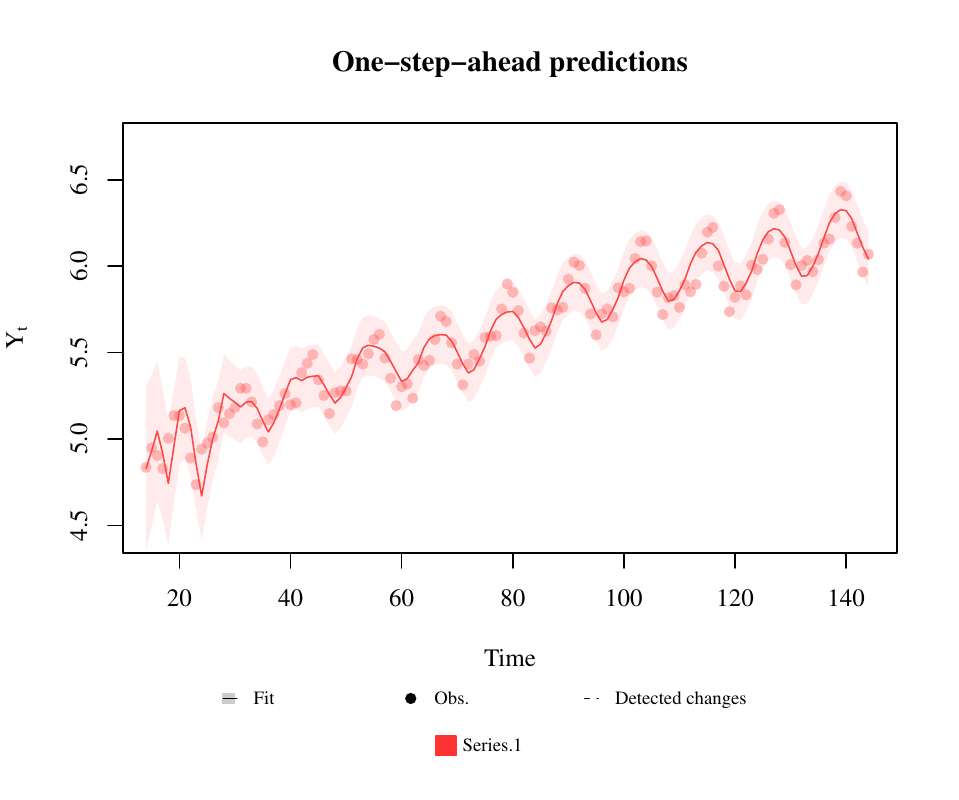}
    \label{fig:plot2}
\end{figure}

Notice that, since this is the univariate case, the \code{data} argument can be a vector.

\subsubsection{Univariated Normal outcome with unknown variance}

For this type of outcome, we assume that:

$$
\begin{aligned}
Y_t|\mu_t,\tau_t &\sim \mathcal{N}\left(\mu_t,\tau_t^{-1}\right),\\
\mu_{t}&=\lambda_{1t},\\
\ln\{\tau_{t}\}&=\lambda_{2t}.\\
\end{aligned}
$$

To create an outcome for this model, we also make use of the \code{Normal} function:

\begin{lstlisting}
Normal(mu, V = NA, Tau = NA, Sd = NA, data)
\end{lstlisting}

Just as before, the \code{mu} argument must be a character representing the label of the linear predictor associated with $\mu_t$. The user must also specify one (and only one) of \code{V}, \code{Tau} or \code{Sd}, which must be a character string representing the label of the associated linear predictor.

Similar to the known variance case, we allow multiple parametrizations of the observational variance. Specifically, if the user provides \code{V}, we assume that $\lambda_{2t}=\ln\{\sigma^2_{t}\}=-\ln\{\tau_t\}$; if the user provides \code{Sd}, we assume that $\lambda_{2t}=\ln\{\sigma_{t}\}=-\ln\{\tau_t\}/2$; if the user provides \code{Tau}, then the default parametrization is used, i.e., $\lambda_{2t}=\ln\{\tau_t\}$. 

The \code{data} argument usually is a $T \times 1$ matrix containing the values of $Y_t$ for each observation. In cases where $\vec{Y}_t$ is univariated, we also accept \code{data} as a line vector, in which case we assume that each coordinate of \code{data} represents the observed value at each time. If a value of data is not available (\code{NA}) for a specific time, it is assumed that there was no observation at that time, thus the update step of the filtering algorithm will be skipped at that time. Note that the evolution step will still be performed, such that the predictive distribution for the missing data and the updated distribution for the latent states at that time will still be provided.

Next, we present a brief example for the usage of this outcome. We use some functions described in the previous sections, as well as some functions that will be presented later on. For now, let us focus on the usage of the \code{Normal} function.

\begin{lstlisting}
structure <- polynomial_block(mu = 1, D = 0.95) +
             polynomial_block(V = 1, D = 0.95)

outcome <- Normal(mu = "mu", V = "V", data = cornWheat$corn.log.return[1:500])
fitted.model <- fit_model(structure, outcome)
plot(fitted.model)
\end{lstlisting}

\begin{figure}[H]
    \centering
    \includegraphics[width=\figsize]{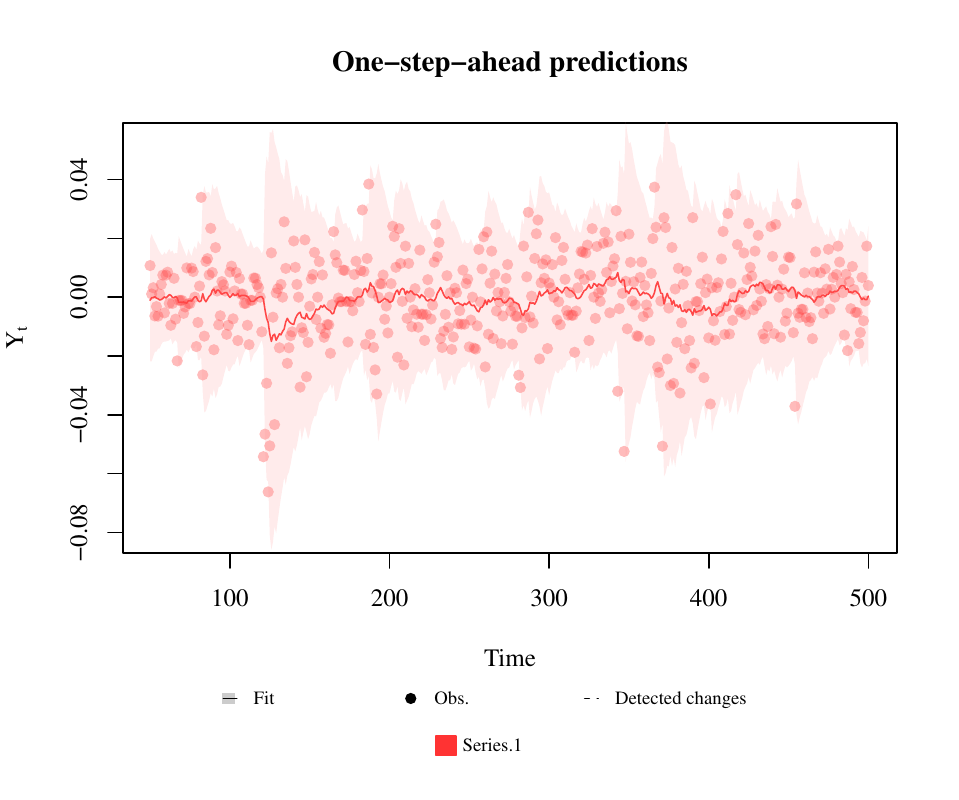}
    \label{fig:plot3}
\end{figure}

Currently, we also support models with bivariate Normal outcomes. In this scenario we assume the following model:

$$
\begin{aligned}
Y_t|\mu_{t},V_t &\sim \mathcal{N}_2\left(\mu_t,V_t\right),\\
\mu_t&=\begin{bmatrix}\mu_{1,t}\\ \mu_{2t}\end{bmatrix},\\
V_t&=\begin{bmatrix}\tau_{1,t}^{-1} & (\tau_{1,t}\tau_{2,t})^{-1/2}\rho_t\\ (\tau_{1,t}\tau_{2,t})^{-1/2}\rho_t & \tau_2^{-1}\end{bmatrix},\\
\mu_{i,t}&=\lambda_{i,t}, i=1,2\\
\tau_{i,t}&=\ln\{\lambda_{(i+2),t}\}, i=1,2\\
\rho_{t}&=\tanh\{\lambda_{5,t}\}.\\
\end{aligned}
$$

Notice that $\rho_t$ represents the \textbf{correlation} (and \textbf{not} the covariance) between the series at time $t$. To guarantee that $\rho_t \in (-1,1)$, we use the Inverse Fisher transformation (also known as the hyperbolic tangent function) as link function.

For those models, \code{mu} must be a character vector, similarly to the case where $V$ is known, and \code{V}, \code{Tau} and \code{Sd} must be a $2 \times 2$ character matrix. The main diagonal elements are interpreted as the linear predictors associated with the precisions, variances or standard deviations, depending if the user used \code{Tau}, \code{V} or \code{Sd}, respectively. The off diagonal elements must be equals (one of them can be \code{NA}) and will be interpreted as the linear predictor associated with $\rho_t$.

Bellow we present an example for the bivariate case:

\begin{lstlisting}
structure <- (polynomial_block(mu = 1, D = 0.95) +
              polynomial_block(log.V = 1, D = 0.95)) * 2 +
             polynomial_block(atanh.rho = 1, D = 0.95)

outcome <- Normal(mu = c("mu.1", "mu.2"),
  V = matrix(c("log.V.1", "atanh.rho", "atanh.rho", "log.V.2"), 2, 2),
  data = cornWheat[1:500, c(4, 5)])
fitted.model <- fit_model(structure, outcome)

##### One-step-ahead predicitons
plot(fitted.model)
\end{lstlisting}

Remember, as seen in Section \ref{Sec:Structure}, that the labels \code{mu.1},\code{mu.2},\code{log.V.1},\code{log.V.2},\code{atanh.rho} are arbitrary and do not affect the model. We recommend the user to use labels that intuitively describe each linear predictor.

\begin{figure}[H]
    \centering
    \includegraphics[width=\figsize]{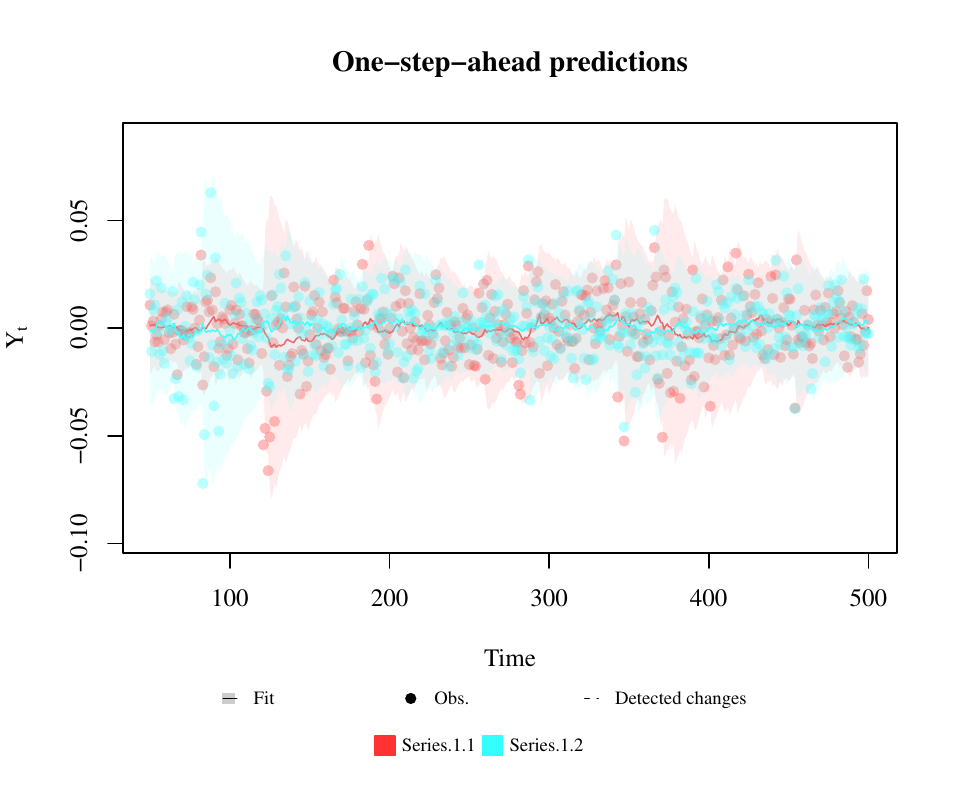}
    \label{fig:plot4}
\end{figure}

\begin{lstlisting}
##### Correlation
plot(coef(fitted.model),'atanh.rho')
\end{lstlisting}

\begin{figure}[H]
    \centering
    \includegraphics[width=\figsize]{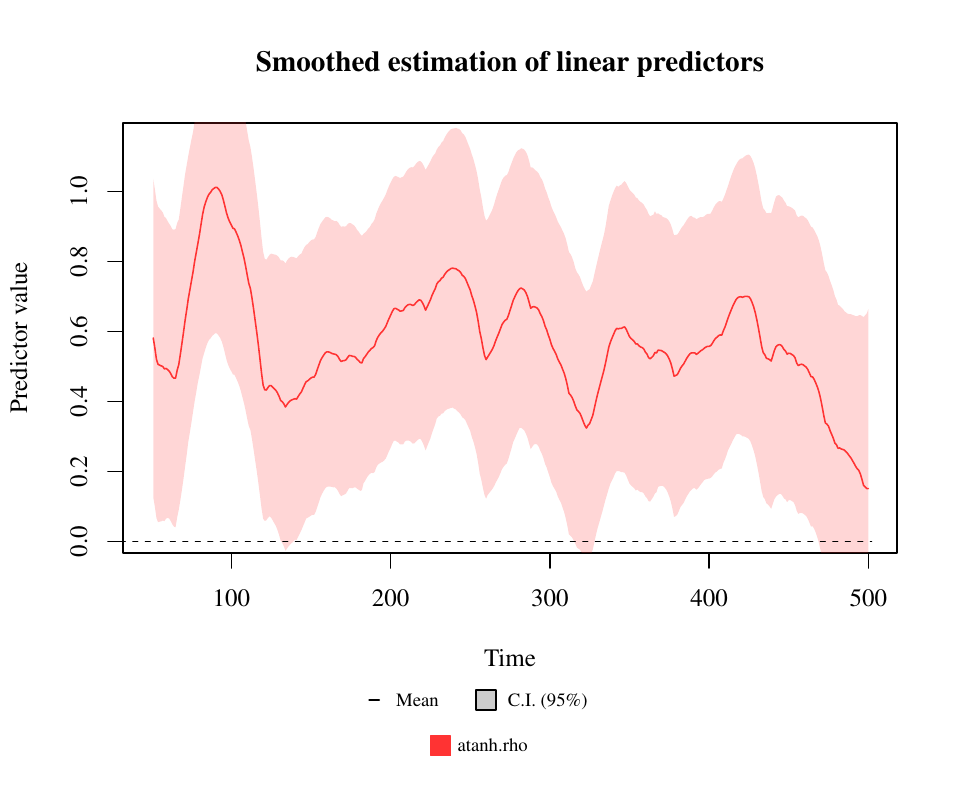}
    \label{fig:plot5}
\end{figure}

Notice that, by the second plot, the correlation between the series (represented by \code{atanh.rho}, i.e., the plot shows $\tanh^{-1}(\rho)$) is significant and changes over time, making the proposed model much more adequate than two independent Normal models (one for each outcome).

\subsection{Handling multiple outcomes}

Lastly, the \pkg{kDGLM} package also allows for the user to jointly fit multiple time series, as long as the marginal distribution of each series is one of the supported distributions \textbf{AND} the series are independent given the latent state vector $\vec{\theta}_t$. In other words, let $\{\vec{Y}_{i,t}\}_{t=1}^{T}, i =1,...,r$, be a set of time series such that:

$$
\begin{aligned}
\vec{Y}_{i,t}|\vec{\eta}_{i,t} &\sim \mathcal{F}_{i}\left(\vec{\eta}_{i,t}\right),\\
g_i(\vec{\eta}_{i,t})&=\vec{\lambda}_{i,t}=F_{i,t}'\vec{\theta}_{t},
\end{aligned}
$$
and $\vec{Y}_{1,t}, ...,\vec{Y}_{r,t}$ are mutually independent given $\vec{\eta}_{1,t}, ...,\vec{\eta}_{r,t}$. Note that the observational distributions $\mathcal{F}_i$ does not need to be the same for each outcome, as long as each $\mathcal{F}_i$ is within the supported marginal distributions. For example, we could have three time series ($r=3$), such that $\mathcal{F}_1$ is a Poisson distribution, $\mathcal{F}_2$ is Normal distribution with unknown mean and precision and $\mathcal{F}_3$ is a Gamma distribution with known shape. Also, this specification does not impose any restriction on the model structure, such that each outcome can have its own component, with polynomial, regression and harmonic blocks, besides having shared components with each other. See \citep{ArtigoMultivar} for a detailed discussion of the approach used to model multiple time series using kDGLMs.

To fit such model, one must only pass the outcomes to the \code{fit_model} function. As an example, we present the code for fitting two Poisson series:

\begin{lstlisting}
structure=polynomial_block(mu.1=1,mu.2=1,order=2,D=0.95)+ # Common factor
          harmonic_block(mu.2=1,period=12,order=2,D=0.975)+ # Seasonality for Series 2
          polynomial_block(mu.2=1,order=1,D=0.95)+ # Local level for Series 2
          noise_block(mu=1)*2 # Overdispersion for both Series

fitted.model=fit_model(structure,
            Adults=Poisson(lambda='mu.1',data=chickenPox[,5]),
            Infants=Poisson(lambda='mu.2',data=chickenPox[,2]))

plot(fitted.model)
\end{lstlisting}

\begin{figure}[H]
    \centering
    \includegraphics[width=\figsize]{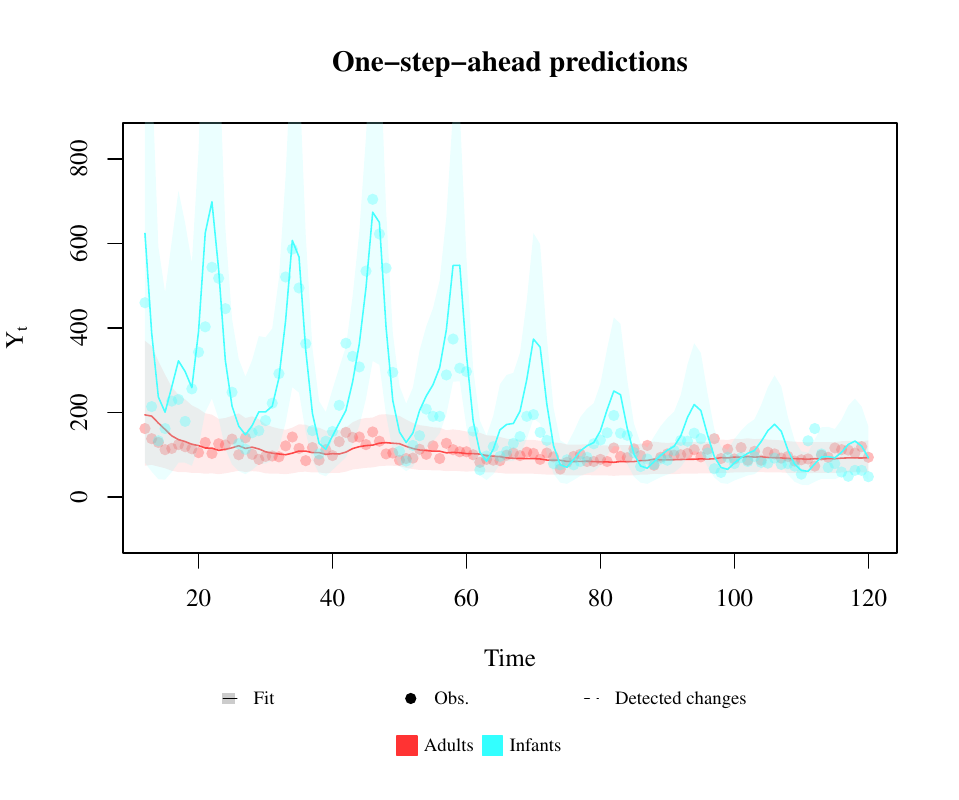}
    \label{fig:plot6}
\end{figure}

It is important to note that the Multivariate Normal and the Multinomial cases are multivariated outcomes and are \textbf{not} considered multiple outcomes on their own, but instead, they are treated as one outcome each, such that the outcome itself is a vector (note that we made no restrictions on the dimension of each $\vec{Y}_{i,t}$). As such, in those cases, the components of the vector $\vec{Y}_{i,t}$ do not have to be mutually independent given $\vec{\eta}_{i,t}$.

Also important to note is that our general approach for modeling multiple time series can not, on its own, be considered a generalization of the Multivariate Normal or Multinomial models. Specifically, if we treat each coordinate of the outcome as a outcome of its own, they would \textbf{not} satisfy the hypotheses of independence given the latent states $\vec{\theta}_t$. This can be compensated with changes to the model structure, but, in general, it is better to model data using a known joint distribution than to assume conditional independence and model the outcomes dependence by shared structure.

\subsubsection{Special case: Conditional modelling}

There is a special type of specification for a model with multiple outcomes that does not require the outcomes to be independent given the latent states. Indeed, if the user specifies the conditional distribution of each outcome given the previous ones, then no hypotheses is needed for fitting the data. 

For instance, lets say that there are three time series $Y_{1,t},Y_{2,t}$ and $Y_{3,t}$, such that each series follows a Poisson distribution with parameter $\eta_{i,t}, i=1,2,3$. Then, $Z_t=Y_{1,t}+Y_{2,t}+Y_{3,t}$ follows a Poisson distribution with parameter $\eta_{1,t}+\eta_{2,t}+\eta_{3,t}$ and $Y_{1,t},Y_{2,t},Y_{3,t}|Z_t$ jointly follows a Multinomial distribution with parameters $N_t=Z_t$ and $\vec{p}_t=\left(\frac{\eta_{1,t}}{\eta_{1,t}+\eta_{2,t}+\eta_{3,t}},\frac{\eta_{2,t}}{\eta_{1,t}+\eta_{2,t}+\eta_{3,t}},\frac{\eta_{3,t}}{\eta_{1,t}+\eta_{2,t}+\eta_{3,t}}\right)'$. Then the user may model $Z_t$ and  $Y_{1,t},Y_{2,t},Y_{3,t}|Z_t$:

\begin{lstlisting}
structure=polynomial_block(mu=1,order=2,D=0.95)+
          harmonic_block(mu=1,period=12,order=2,D=0.975)+
          noise_block(mu=1)+polynomial_block(p=1,D=0.95)*2

outcome1=Poisson(lambda='mu',data=rowSums(chickenPox[,c(2,3,5)]))
outcome2=Multinom(p=c('p.1','p.2'),data=chickenPox[,c(2,3,5)])

fitted.model=fit_model(structure, Total=outcome1, Proportions=outcome2)

plot(fitted.model)
\end{lstlisting}

\begin{figure}[H]
    \centering
    \includegraphics[width=\figsize]{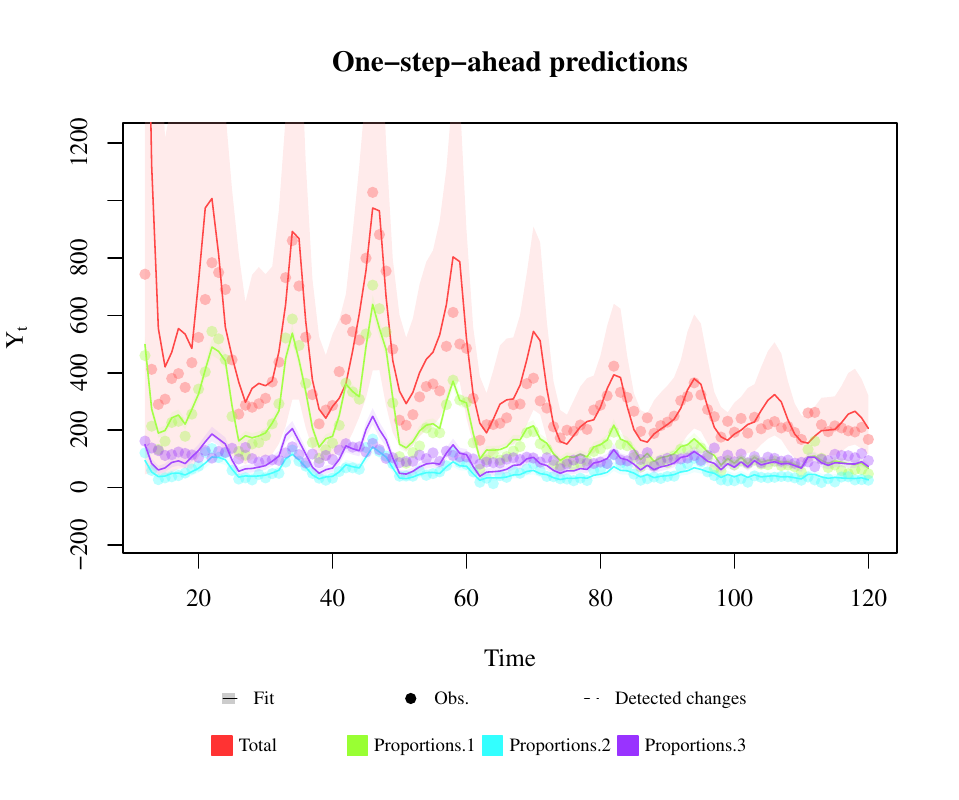}
    \label{fig:plot7}
\end{figure}

See \cite{AlexMP} for a discussion of Multinomial-Poisson models. More applications are presented in the advanced examples section of the vignette.

\section{Fitting and analysing models}\label{Sec:Fitting}

In this section we briefly present the usage of the \code{fit_model} function, along side the auxiliary functions for analyzing a fitted model, such as the \code{summary}, \code{coef},\code{plot} and \code{forecast} methods, and the \code{search_grid} and \code{dlm_sampling} functions. For a deep dive in the details of each argument of each function, see the documentation of those function and/or the reference manual.

\subsection{Filtering and smoothing}
 
The usage of the \code{fit_model} function is as follows:

\begin{lstlisting}
fit_model(..., smooth = TRUE, p_monit = NA)
\end{lstlisting}

The \code{...} argument receives \code{dlm_block} and \code{dlm_distr} objects, the creation of which was described in the previous sections. In particular, if the user gives a \code{dlm_distr} object as a named argument, its name is used as the label for that outcome.

The argument \code{smooth} is a Boolean indicating if the smoothed distribution of the latent states should be evaluated. In general, we recommend the users to not change this value, as the computational cost of smoothing is usually negligible. 

\code{p_monit} controls the sensibility of the automated monitoring and we shall discuss its usage in subsection \ref{intervention}.

Bellow we present a code that fits a Poisson model:

\begin{lstlisting}
level <- polynomial_block(rate = 1, order = 2, D = 0.95)
season <- harmonic_block(rate = 1, period = 12, order=2, D = 0.975)

outcome <- Poisson(lambda = "rate", data = c(AirPassengers))

fitted.model <- fit_model(
  level, season,           # Strucuture
  AirPassengers = outcome) # outcome
\end{lstlisting}

The first two lines create structural blocks representing a random walk on $\mu_t$ and a seasonal component described by harmonics. The fourth line creates a Poisson outcome such that the rate parameter \code{lambda} is equal to $\exp\{\text{rate}\}$, where \code{rate} is the label given to the linear predictor when creating the structural blocks (see section \ref{Sec:Structure} for details about the linear predictor). The last line receives the model structure and the Poisson outcome and fits the model, obtaining the parameters for the filtered and smoothed distribution of all latent states.

The user can see how the model fits the data using the \code{plot} method, the syntax of which is as follows:

\begin{lstlisting}
plot.fitte_dlm(model, pred.cred = 0.95, lag = 1, cutoff = floor(model$t/10))
\end{lstlisting}

The \code{model} argument must be a \code{fitted_dlm} object (i.e., the output of the \code{fit_model} function). 

\code{pred.cred} must be a number between $0$ and $1$ representing the desired credibility of the predictions.

\code{lag} must be an integer representing the number of steps ahead to be used for predictions. If \code{lag}$<0$, the smoothed distribution is used for predictions and, if \code{lag}$=0$, the filtered distribution is used Instead.

\code{cutoff} must be an integer representing the number of initial steps that should be omitted in the plot. Usually, the model is still learning in the initial steps, so the predictions are not reliable. The default value is one tenth of the sample size rounded down.

Lastly \code{plot.pkg} must be a string specifying the plot engine to be used. Should be one of \code{'auto'}, \code{'base'}, \code{'ggplot2'} or \code{'plotly'}. If \code{'auto'} is used, then the function tries to use the \pkg{plotly} package, if it fails, then it tries to use the \pkg{ggplot2} packge and, if it also fails, the native functions provided by \proglang{R} will be used.

\begin{lstlisting}
plot(fitted.model)
\end{lstlisting}

\begin{figure}[H]
    \centering
    \includegraphics[width=\figsize]{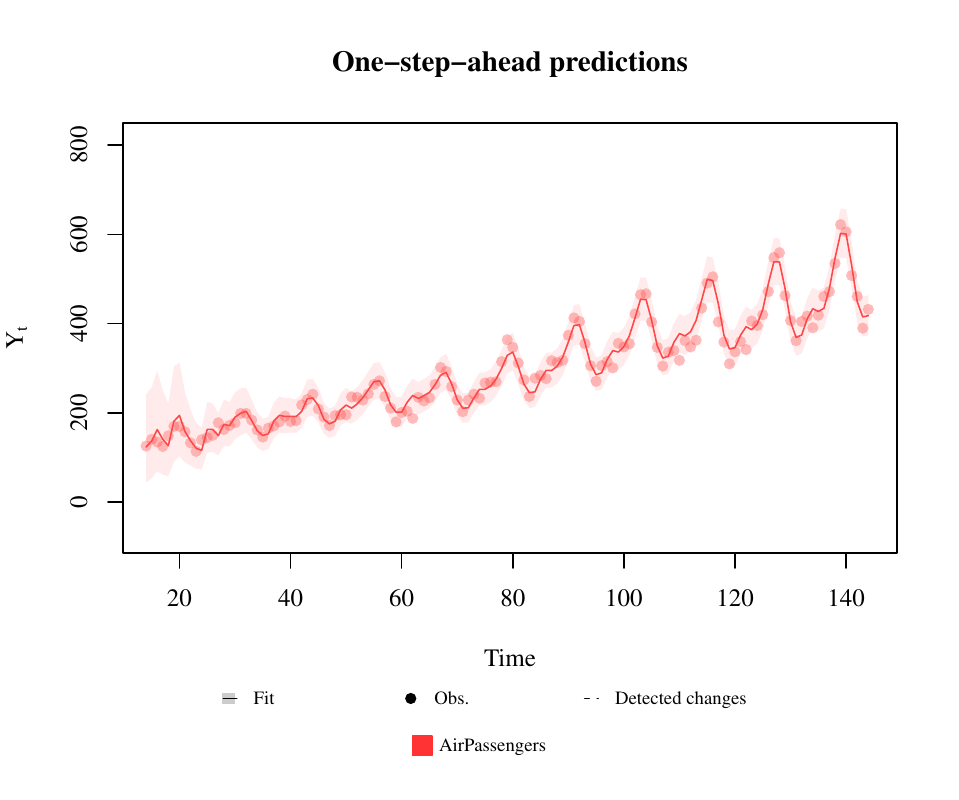}
    \label{fig:plot8}
\end{figure}

The remaining functions and methods in this section have similar usage as the \code{plot.fitte_dlm} method, as such, for the sake of brevity, we will only highlight the unique arguments and/or behaviors of each function or method present. We strongly advise the user to consult the reference manual and the documentation of each function for detailed descriptions of any function. 

To see a summary of the fitted model, one can use the \code{summary} method:

\begin{lstlisting}
summary(fitted.model)
\end{lstlisting}

\begin{\verbatimfont} \begin{verbatim}
Fitted DGLM with 1 outcomes.

distributions:
    AirPassengers: Poisson

Coeficients (smoothed) at time 144:
                Estimate Std. Error   t value Pr(>|t|)
Var.Poly.Level   6.20799    0.01413 439.36382   <1e-12 *** 
Var.Poly.Slope   0.00889    0.00052  17.05167   <1e-12 *** 
Var.Sazo.Main.1 -0.15753    0.01196 -13.17551   <1e-12 *** 
Var.Sazo.Aux.1  -0.07526    0.01225  -6.14283 8.11e-10 *** 
Var.Sazo.Main.2 -0.00841    0.01167  -0.72095    0.471   
Var.Sazo.Aux.2   0.08618    0.01197   7.19746   <1e-12 *** 
---
Signif. codes:  0 '***' 0.001 '**' 0.01 '*' 0.05 '.' 0.1 ' ' 1

---
One-step-ahead prediction
Log-likelihood        : -567.01085
Interval Score        :  102.96154
Mean Abs. Scaled Error:    0.43384
Relative abs. Error   :    0.05214
Mean Abs. Error       :   13.89588
Mean Squared Error    :  320.56218
---
 \end{verbatim}\end{\verbatimfont}

Notice that the coefficients shown are those of the last observation, but one can also see the summary of any particular time by changing the \code{t} argument:

\begin{lstlisting}
summary(fitted.model, t = 100)
\end{lstlisting}

\begin{\verbatimfont} \begin{verbatim}
Fitted DGLM with 1 outcomes.

distributions:
    AirPassengers: Poisson

Coeficients (smoothed) at time 100:
                Estimate Std. Error   t value Pr(>|t|)
Var.Poly.Level   5.84815    0.01057 553.42205   <1e-12 *** 
Var.Poly.Slope   0.01003    0.00082  12.24626   <1e-12 *** 
Var.Sazo.Main.1  0.01732    0.01032   1.67782    0.093   
Var.Sazo.Aux.1   0.16878    0.01009  16.72997   <1e-12 *** 
Var.Sazo.Main.2 -0.06611    0.00995  -6.64338 3.07e-11 *** 
Var.Sazo.Aux.2  -0.05225    0.00999  -5.23234 1.67e-07 *** 
---
Signif. codes:  0 '***' 0.001 '**' 0.01 '*' 0.05 '.' 0.1 ' ' 1

---
One-step-ahead prediction
Log-likelihood        : -567.01085
Interval Score        :  102.96154
Mean Abs. Scaled Error:    0.43384
Relative abs. Error   :    0.05214
Mean Abs. Error       :   13.89588
Mean Squared Error    :  320.56218
---
 \end{verbatim}\end{\verbatimfont}

For more details about the usage of the \code{summary} method, see the associated documentation (\code{help(summary.fitted_dlm)}).

\subsection{Extracting components}

Naturally, the user may want to extract information about the posterior distribution of the parameters of the fitted model, such that a more thorough analysis may be performed. For extracting the parameters of the distribution of latent states, linear predictors and observational model parameters, one can use the \code{coef} method:

\begin{lstlisting}
coef(object, eval_t = seq_len(object$t), lag = -1, pred.cred = 0.95, eval.pred = FALSE, eval.metric = FALSE, ...)
\end{lstlisting}

The \code{object} parameter must be specified as a \code{fitted_dlm} object, which represents the model from which the components will be extracted. The \code{eval_t} parameter should be a vector that denotes the time indices at which the posterior is to be evaluated. The parameters \code{lag} and \code{pred.cred} retain their meanings analogous to those in the plot method of the \code{fitted_dlm} class. The \code{eval.pred} parameter requires a boolean value, indicating whether the predictive distribution for the observations is to be evaluated. Additionally, the \code{eval.pred} parameter should be a boolean specifying whether the model comparison metrics are to be computed.

The output of this function is a \code{dlm_coef} object containing:

\begin{itemize}
\item  \code{data}: A data frame with the model evaluated at each observed time.

\item  \code{mt}: A $n \times T$ matrix representing the mean of the latent states at each time, where $n$ is the number of latent states in the model and $T$ is the length of the time series;

\item  \code{Ct}: A 3D-array containing the covariance matrix of the latent state at each time. Dimensions are $n \times n \times T$;

\item  \code{ft}: A $k \times T$ matrix representing the mean of the linear predictor at each time, where $k$ is the number of linear predictors in the model and $T$ is the length of the time series;

\item  \code{Qt}: A 3D-array containing the covariance matrix for the linear predictor at each time. Dimensions are $k \times k \times T$;

\item  Several vectors with some metrics, including the \textbf{Mean Absolute Error} (MAE), the \textbf{Mean Absolute Scaled Error} (MASE) \citep{mase}, the \textbf{Relative Absolute Error} (RAE), the \textbf{Mean Squared Error} (MSE) and the \textbf{Interval Score} (interval.score) \citep{interval_score}.

\item  \code{conj.param}: A list containing the conjugated parameters for the distribution of the parameter of the observational model.
\end{itemize}

It is important to highlight that, following the method proposed in \cite{ArtigokParametrico}, the joint distribution of the latent states and linear predictors at each time is Gaussian, such that the mean vector and covariance matrix completely define their distribution.

The user may also want to plot the latent states, for which the \code{plot} method for the \code{dlm_coef} class can be used:

\begin{lstlisting}
plot(fitted.coef)
\end{lstlisting}

\begin{figure}[H]
    \centering
    \includegraphics[width=\figsize]{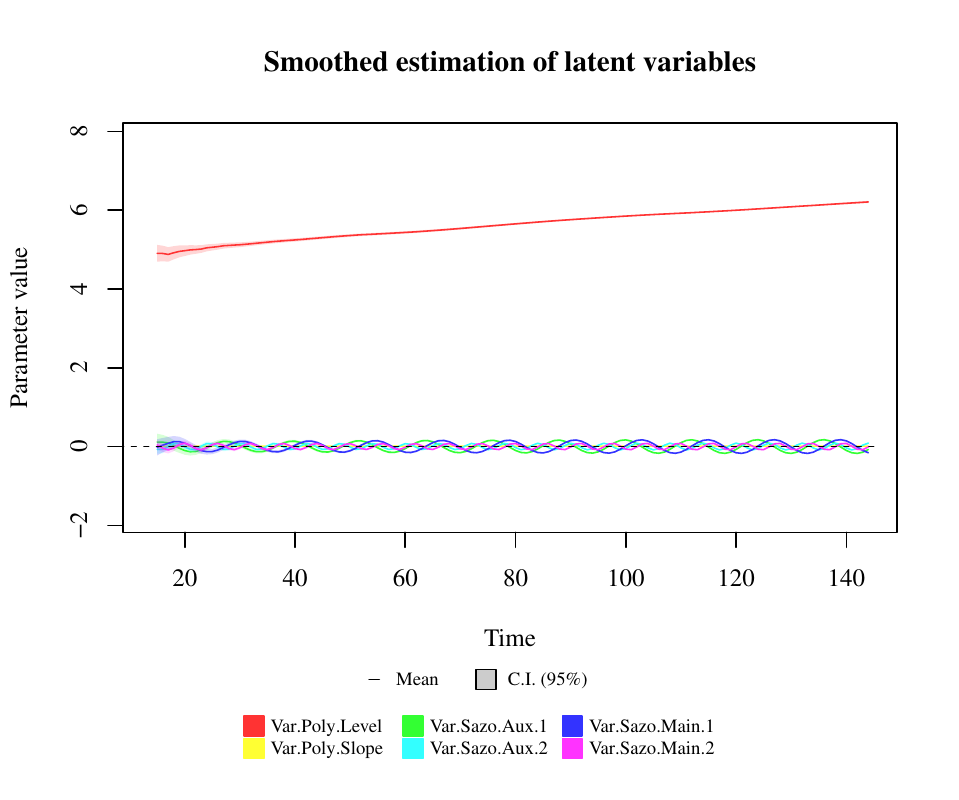}
    \label{fig:plot9}
\end{figure}

If the user wants to see only a restricted set of latent states, the extra argument \code{var} can be used to specify the label of the variables to plot:

\begin{lstlisting}
plot(fitted.coef,'Var.Poly.Level')
\end{lstlisting}

\begin{figure}[H]
    \centering
    \includegraphics[width=\figsize]{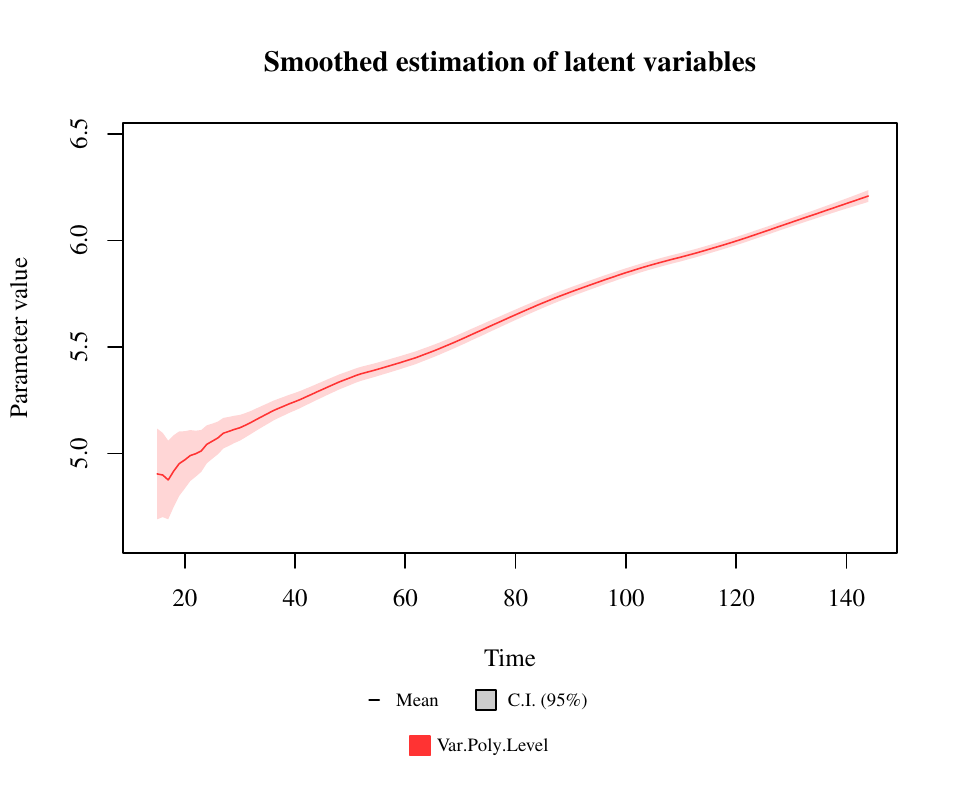}
    \label{fig:plot10}
\end{figure}

The user may also plot the linear predictors, by specifying the name of the linear predictor:

\begin{lstlisting}
plot(fitted.coef,'rate')
\end{lstlisting}

\begin{figure}[H]
    \centering
    \includegraphics[width=\figsize]{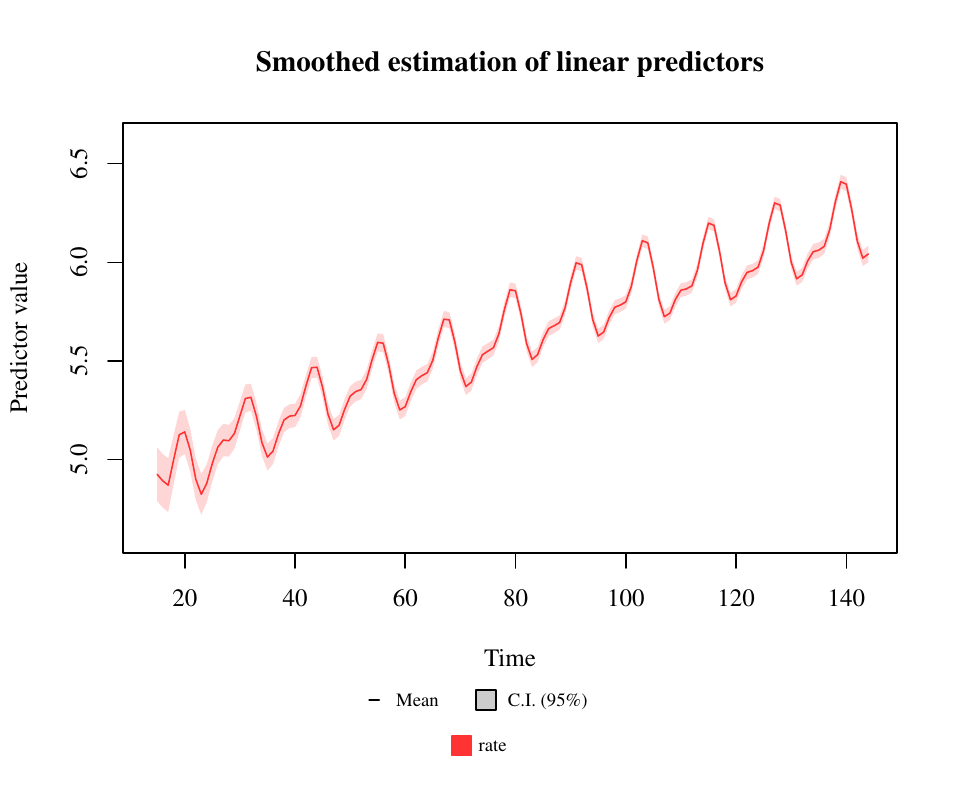}
    \label{fig:plot11}
\end{figure}

Lastly, although we do not recommend it, the user may also extract some of these information directly from the \code{fitted_dlm} object.

We strongly recommend every user to consult the documentation of each of these functions to see the full set of features provided by the \pkg{kDGLM} package.

\subsection{Forecasting}

Notice that all methods and functions presented previously were restricted to the period where the model was fitted. If the user wishes make predictions for future observations, the \code{forecast} method can be used:

\begin{lstlisting}
forecast(object, t = 1,
         plot = ifelse(requireNamespace("plotly", quietly = TRUE), "plotly", ifelse(requireNamespace("ggplot2", quietly = TRUE), "ggplot2", "base")),
         pred.cred = 0.95,
         ...)
\end{lstlisting}

The \code{object} parameter is required to be a \code{fitted_dlm} object. The \code{t} parameter specifies the prediction window. The \code{plot} parameter determines whether a plot should be generated and, if applicable, which engine to use, akin to the plot method in the \code{fitted_dlm} class. The \code{pred.cred} parameter signifies the credibility of the confidence intervals.

\begin{lstlisting}
forecast(fitted.model, t = 20)$plot
\end{lstlisting}

\begin{figure}[H]
    \centering
    \includegraphics[width=\figsize]{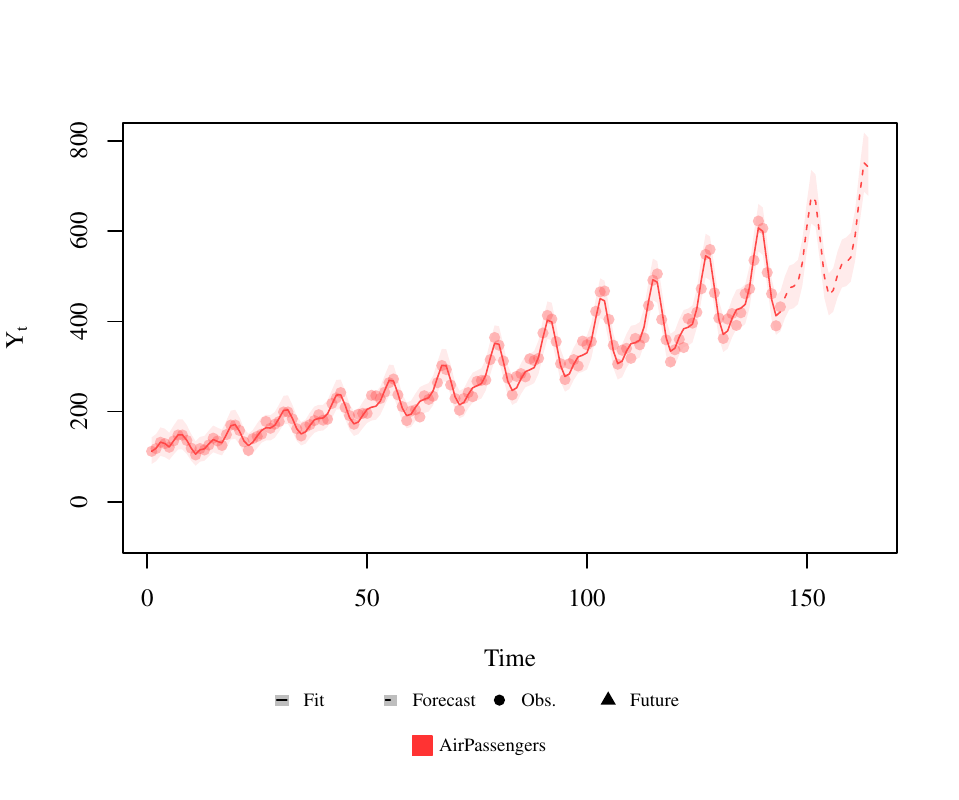}
    \label{fig:plot12}
\end{figure}

Additionally to a plot (which is optional), the \code{forecast} method for the \code{fitted_dlm} class also provides a similar set of attributes of that which the \code{dlm_coef} class has, specifically, the predictive distribution for the latent states, the linear predictors and the observational model parameters, along with the predictions for future observations.

It is relevant to point out that if external data is necessary for forecasting, such as for models with regressive blocks or transfer functions, it is necessary to pass those values for the \code{forecast} method. In this scenario, the user must pass a new argument named as the variable that is "missing" from the model. See the documentation to see how to determine the name of the missing values or, more practically, try to use the \code{forecast} method without the necessary arguments, since the name of the missing variables will be presented in the error message.

Here we present two examples for a model with Multinomial outcome: One where the covariates where not properly passed and another where they were:

\begin{lstlisting}
 structure <-
   polynomial_block(p = 1, order = 2, D = 0.95) +
   harmonic_block(p = 1, period = 12, D = 0.975) +
   noise_block(p = 1, R1 = 0.1) +
   regression_block(
     p = chickenPox$date >= as.Date("2013-09-1"),
     # Vaccine was introduced in September of 2013
     name = "Vaccine")

 outcome <- Multinom(p = c("p.1", "p.2"), data = chickenPox[, c(2, 3,  5)])
 fitted.model <- fit_model(structure * 2, chickenPox = outcome)

 forecast(fitted.model, t = 24) # Missing extra arguments
\end{lstlisting}

\begin{\verbatimfont} \begin{verbatim}
Error in  forecast(fitted.model, t = 24)  : 
  Error: Missing extra arguments: Vaccine.1.Covariate, Vaccine.2.Covariate
 \end{verbatim}\end{\verbatimfont}

\begin{lstlisting}
 forecast(fitted.model, t=24,
   Vaccine.1.Covariate = rep(TRUE, 24), # Extra argument for covariate 1
   Vaccine.2.Covariate = rep(TRUE, 24))  # Extra argument for covariate 2
\end{lstlisting}

\begin{figure}[H]
    \centering
    \includegraphics[width=\figsize]{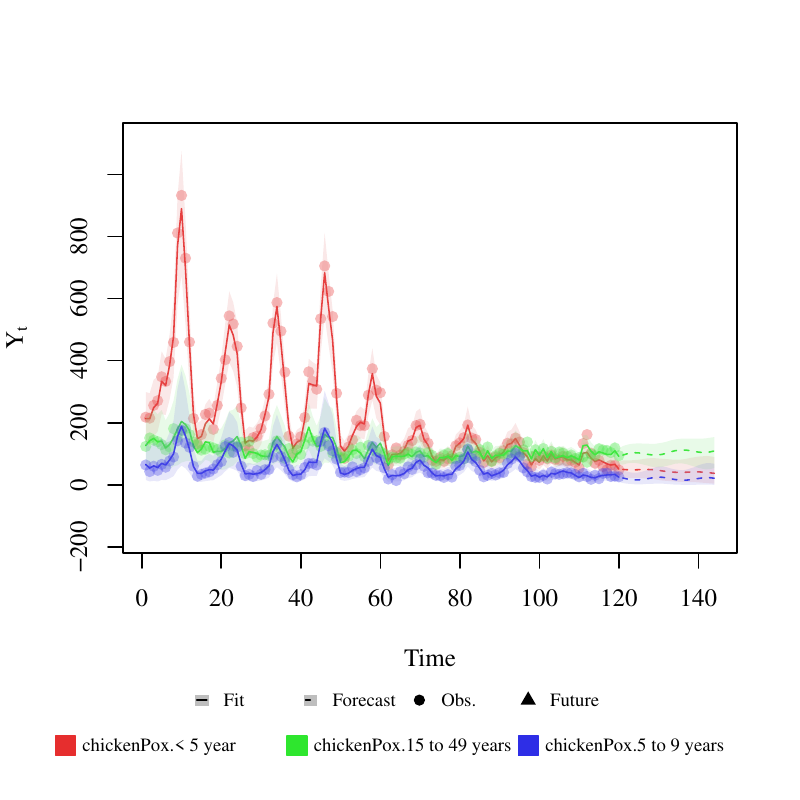}
    \label{fig:plot26}
\end{figure}

For more details on the usage of this function, see the associated documentation.

\subsection{Updating a fitted model}

One of the major advantages of the sequential nature of the methodology proposed in \cite{ArtigokParametrico} is that it allows for the updating of the posterior distribution of the parameter when new data arrives, but without the necessity of reprocessing the data previously observed. This feature is particularly useful in problems that involve monitoring or real time inference about a phenomena.

For updating a \code{fitted_dlm} object, the user can use the \code{update} method for the \code{fitted_dlm} class:

\begin{lstlisting}
update.fitted_dlm(object, ...)
\end{lstlisting}

The \code{object} argument must be a \code{fitted_dlm} object. Moreover, the \code{...} argument must be a sequence of named arguments containing the new information observed. For example:

\begin{lstlisting}
level <- polynomial_block(rate = 1, order = 2, D = 0.95)
season <- harmonic_block(rate = 1, period = 12, order=2, D = 0.975)
# Omitting the last 44 observations
outcome <- Poisson(lambda = "rate", data = c(AirPassengers)[1:100])
fitted.model <- fit_model(
  level, season,           # Strucuture
  AirPassengers = outcome) # outcome
updated.fit=update(fitted.model,
                   AirPassengers=list(data=c(AirPassengers)[101:144]))
\end{lstlisting}

Note that the name of the argument containing the new observations must be the label given to that outcome when first fitting the model. In this case, the argument must be named \code{AirPassengers}, as this was the label used in the \code{fit_model} function. If a label was not provided when fitting the model, a default name will be used, which consist of the string \code{'Series.'} followed by a proper index for that outcome. 

The \code{update} function may require extra arguments containing covariates, pulses (for the transfer function), the offset, etc.. In such cases, the syntax is the same as the \code{forecast} method.
 
\subsection{Intervention and monitoring}\label{intervention}

As a key feature, the \pkg{kDGLM} package has support for intervention and automated monitoring. First, if the user is aware that at some specific time there is some change in the time series that is not part of its temporal dynamic, then the user should provide that information in the model structure. For that we provide the \code{intervention} function:

\begin{lstlisting}
data <- c(AirPassengers)
# Adding an artificial change, so that we can make an intervention on the data at that point
# Obviously, one should NOT change their own data.
data[60:144] <- data[60:144] + 100

level <- polynomial_block(rate = 1, order = 2, D = 0.95)
season <- harmonic_block(rate = 1, order = 2, period = 12, D = 0.975)

# Reducing the discount factor so that the model can capture the expected change.
level <- level |> intervention(time = 60, D = 0.005, var.index = 1)
# Comment the line above to see the fit without the intervention

fitted.model <- fit_model(level, season,
  AirPassengers = Poisson(lambda = "rate", data = data))

plot(fitted.model)
\end{lstlisting}

\begin{figure}[H]
    \centering
    \includegraphics[width=\figsize]{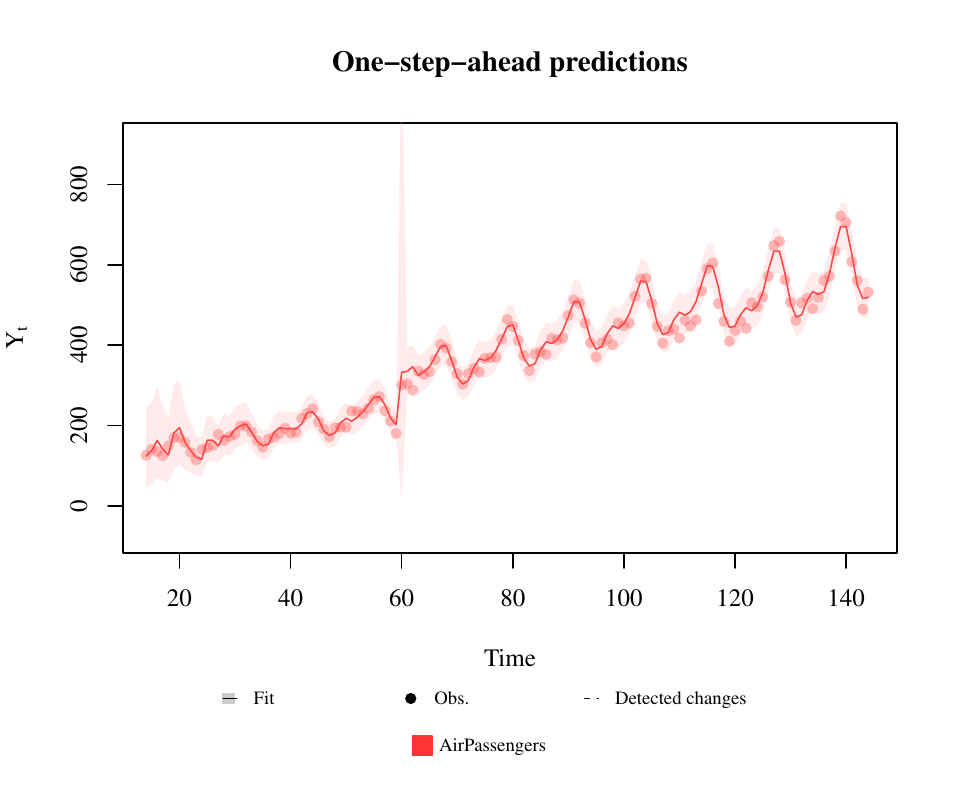}
    \label{fig:plot13}
\end{figure}

See the documentation of the \code{intervention} function for more details about its arguments. Also, we strongly recommend the user to consult \cite{WestHarr-DLM}, chapter 11 for a detailed discussion about Feed-Foward Interventions.

In case the user is not aware of any behavioral changes in the data, but suspects that they may have occurred at some unknown time, then we recommend the use of automated monitoring.

To fit a model using automated monitoring, the user must provide a valid value for the \code{p.monit} argument in the \code{fit_model} function. This argument receives values between $0$ and $1$, such that its value is interpreted as the prior probability (i.e., the probability before observing the data), at any given time, of behavioral change in the series that is not accommodated by the temporal dynamic.

\begin{lstlisting}
data <- c(AirPassengers)
# Adding an artificial change, so that we can make an intervention on the data at that point
# Obviously, one should NOT change their own data.
data[60:144] <- data[60:144] + 100

level <- polynomial_block(rate = 1, order = 2, D = 0.95)
season <- harmonic_block(rate = 1, order = 2, period = 12, D = 0.975)

fitted.model <- fit_model(level, season,
  AirPassengers = Poisson(lambda = "rate", data = data), 
  p.monit=0.05)

plot(fitted.model)
\end{lstlisting}
\begin{figure}[H]
    \centering
    \includegraphics[width=\figsize]{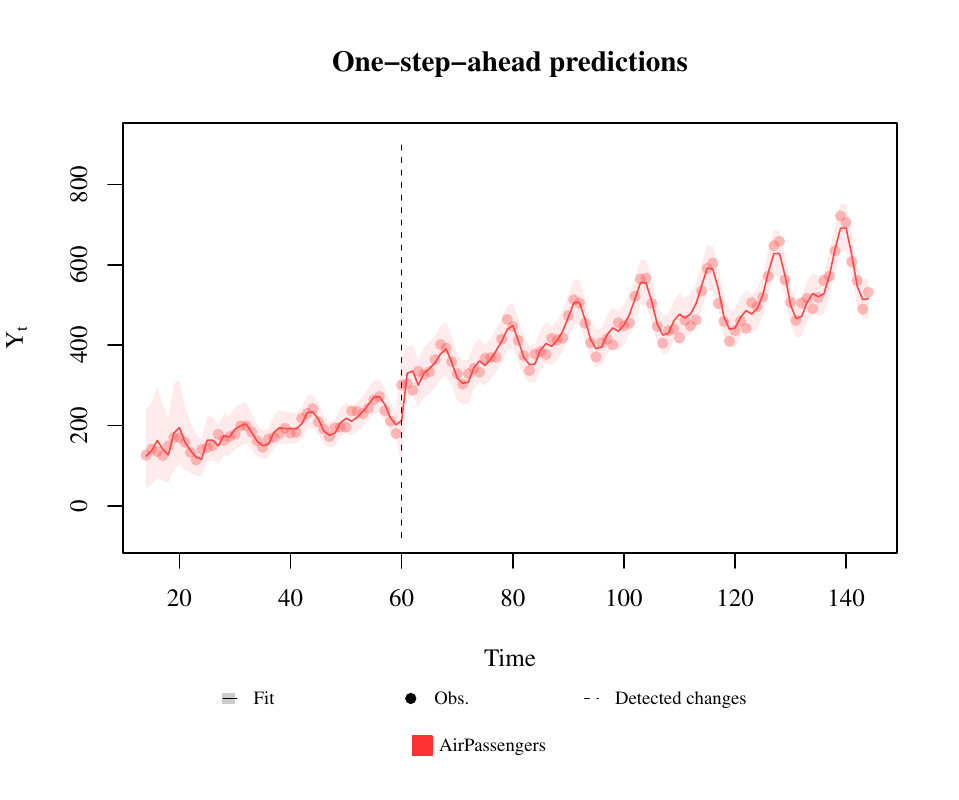}
    \label{fig:plot14}
\end{figure}

The approach used for automated monitoring is almost identical to that of \cite{WestHarr-DLM}, chapter 11.4, using Bayes' factor, such that \code{p.monit}$=0.05$ yields a threshold equivalent to that recommended in \cite{WestHarr-DLM}.

\subsection{Tools for sensibility analysis}\label{Subsec:sense_analysis}

In some situations, the user may not be sure about which value to use for some hyperparameter of the model (such as the discount factor or the order of a block) or about the inclusion of some structural block. As such, one might choose to perform a sensibility analysis on the effect of those choices. For such analysis, we provide the \code{search_model} function.

As an motivational example, let us assume that we are unsure about which value to choose for the discount factor in a polynomial trend block of a Poisson model. First, when defining the model structure, we must set the discount factor as a string, which will be used as the label for the unspecified parameter:

\begin{lstlisting}
level <- polynomial_block(rate = 1, order = 2, D = 'D1')
\end{lstlisting}

By setting the discount factor as a string, the structural block becomes partially \textbf{undefined}:

\begin{lstlisting}
summary(level)
\end{lstlisting}

\begin{\verbatimfont} \begin{verbatim}
Basic DLM block.
Latent states: 
    Var.Poly: Level, Slope (2 variable(s))

Linear predictors: 
    rate

Status: undefined
Serie length: 1
Interventions at: 
Number of latent states: 2
Number of linear predictors: 1
 \end{verbatim}\end{\verbatimfont}

As such, this block \textbf{can not} be used in the \code{fit_model} function:

\begin{lstlisting}
season <- harmonic_block(rate = 1, order = 2, period = 12, D = 0.975)
outcome <- Poisson(lambda = "rate", data = c(AirPassengers))
fitted.model <- fit_model(level, season, AirPassengers = outcome)
\end{lstlisting}

\begin{\verbatimfont} \begin{verbatim}
Error in fit_model(level, season, AirPassengers = outcome) : 
  Error: One or more hiper parameter are undefined.
  Did you meant to use the search_model function?
 \end{verbatim}\end{\verbatimfont}

As the user can see in the error message above, an undefined \code{dlm_block} can only be used with the \code{search_model} function. This function is used to fit a set of models, while computing some comparative metrics.

\begin{lstlisting}
search_model(..., search.grid, condition = "TRUE", 
             metric = "log.like", smooth = TRUE, lag = 1, 
             pred.cred = 0.95, metric.cutoff = NA,
             p.monit = NA)
\end{lstlisting}

The usage of the \code{search_model} function is almost identical to that of the \code{fit_model} function, the \code{...}, \code{smooth} and \code{p.monit} arguments having the exact same meaning.

The \code{search.grid} argument must be a list containing, for each undefined parameter, the list of values to be tested. By default, this function will test all possible combinations of the undefined parameter. If the user wishes to skip some combinations, the \code{condition} argument can be used to provide a string with the criterion to determine which combinations shall be evaluated.

The remaining options provide some control over the comparative metrics. The \code{metric} argument (\code{'mae'}, \code{'mase'}, \code{'rae'}, \code{'log.like'} or \code{'interval.score'}) indicates which metric to use when selecting the best model (all metrics are calculated, not matter the value of the \code{metric} argument, but only the best model by the chosen metric is saved). The \code{lag} argument indicates the number of steps ahead to be used for predictions ($0$ indicates filtered predictions and negative values indicate smoothed predictions). The \code{pred.cred} argument indicates the credibility of the intervals used when computing the Interval Score. The \code{metric.cutoff} argument indicates the number of initial observations to be ignored when computing the metrics.

After evaluating all valid combinations of hyper parameters, the \code{search_model} function returns the best model by the chosen metric, along with a data frame containing the metrics for each model evaluated.

\begin{lstlisting}
level <- polynomial_block(rate = 1, order = 2, D = 'D.level')
season <- harmonic_block(rate = 'sazo.effect', period = 12,
                         order = 2, D = "D.sazo")

outcome <- Poisson(lambda = "rate", data = c(AirPassengers))

search_model(level, season,outcome,
  search.grid = list(
    sazo.effect = c(0, 1),
    D.level = seq(0.8, 1, l = 11),
    D.sazo = seq(0.95, 1, l = 11)),
  condition = "sazo.effect==1 | D.sazo==1"
)$search.data |> head()
\end{lstlisting}

It is important to note that not all hyper parameters can be tested directly by the search model function, indeed, only the components associated with $F_t$, $D_t$, $h_t$, $H_t$, $a_1$ and $R_1$ can be treated as undefined. Still, if the user wants to test some other hyper parameter that cannot be tested directly (such as the order of a polynomial block or the period of a harmonic block), he can create one block for each option and perform a sensibility analysis for the inclusion/exclusion of each block:

\begin{lstlisting}
# Creating a block for each order
level <- polynomial_block(rate = 'pol.ord.1', order = 1, D = 0.95)+
         polynomial_block(rate = 'pol.ord.2', order = 2, D = 0.95)+
         polynomial_block(rate = 'pol.ord.3', order = 3, D = 0.95)+
         polynomial_block(rate = 'pol.ord.4', order = 4, D = 0.95)
season <- harmonic_block(rate = 1, order = 2, period = 12, D = 0.975)

outcome <- Poisson(lambda = "rate", data = c(AirPassengers))

search_model(level, season,outcome,
  search.grid = list(
    # Each block can be present (1) or absent (0).
    pol.ord.1 = c(0, 1), pol.ord.2 = c(0, 1),
    pol.ord.3 = c(0, 1), pol.ord.4 = c(0, 1)),
  condition = "pol.ord.1+pol.ord.2+pol.ord.3+pol.ord.4==1"
  # Only test combinations with exactly one polynomial block.
)$search.data |> head()
\end{lstlisting}

\subsection{Sampling and hyper parameter estimation}

Lastly, one may also want to draw samples from the latent states, linear predictors and/or the parameters $\eta_t$ of the observational model. This can be useful to evaluate non-linear functions of the model parameters or when the DGLM is only a part of a bigger model, from which the parameters are being estimated with Gibbs Algorithm. It is important to note that, with the method proposed in \cite{ArtigokParametrico}, sampling from the posterior distribution of the latent states is straight forward, allowing the user to obtain large independent (\textbf{not} approximately independent, but \textbf{exactly} independent) samples with very low computational cost. See \cite{WestHarr-DLM}, chapter 15, for details about the sampling algorithm.

The \pkg{kDGLM} package offers the \code{dlm_sampling} function, which provides a routine for drawing independent samples from any fitted model:

\begin{lstlisting}
dlm_sampling(fitted.model, 5000)
\end{lstlisting}

For the example above, where our model has $6$ latent states and $144$ observations (which yields a total of $864$ parameters), it takes approximately $0.3$ seconds to draw a sample of size $5.000$.

Another useful feature of the \pkg{kDGLM} package is that it provides an approximate value for the Model likelihood $f(y)= \int_{\mathbb{R}^{n}}f(y|\theta)f(\theta)d\theta$, where $y$ represents the values for $Y_t$ for all $t$ and $\theta$ represents the values of $\theta_t$ for all $t$. This feature can be used for two main purposes: to compare different models and to evaluate the posterior distribution of hyper parameter. 

To compare different models, $\mathcal{M}_1,...,\mathcal{M}_k$ , one can note that $f(\mathcal{M}_i|y) \propto f(y|\mathcal{M}_i)f(\mathcal{M}_i)$, where $f(y|\mathcal{M}_i)= \int_{\mathbb{R}^{n}}f(y|\theta,\mathcal{M}_i)f(\theta|\mathcal{M}_i)d\theta$ is the likelihood of model $\mathcal{M}_i$ and $f(\mathcal{M}_i)$ is the prior for model $\mathcal{M}_i$. To evaluate $f(y|\mathcal{M}_i)$, one can make use of the \code{coef} method, the \code{eval_dlm_norm_const} function or \code{search_model} function by setting \code{lag} to a negative number (the log likelihood metric will be $\ln f(y|\mathcal{M}_i)$). Similarly, if the user wants to obtain the marginal posterior distribution of an hyper parameter $\tau$, it can observe that $f(\tau|y) \propto f(y|\tau)f(\tau)$, from which $f(y|\tau)$ can be evaluated using the \code{coef} method, the \code{eval_dlm_norm_const} function or \code{search_model} function.

\begin{lstlisting}
H.range=seq.int(0,1,l=100)
log.like.H=seq_along(H.range)
log.prior.H=dlnorm(H.range,0,1,log=TRUE)
for(i in seq_along(H.range)){
  level <- polynomial_block(rate = 1, order = 2, H=H.range[i])
  season <- harmonic_block(rate = 1, order = 2, period = 12, D = 0.975)
  # Using only 10 observations, for the sake of a pretty plot. For this particular application, the posterior density of H rapidly becomes highly consentrated in a single value.
  outcome <- Poisson(lambda = "rate", data = c(AirPassengers)[1:10]) 
  fitted.model <- fit_model(level,season,outcome)
  log.like.H[i]=eval_dlm_norm_const(fitted.model)
}
log.post.H=log.prior.H+log.like.H
post.H=exp(log.post.H-max(log.post.H))
plot(H.range,post.H,type='l',xlab='H',ylab='f(H|y)')
\end{lstlisting}

\begin{figure}[H]
    \centering
    \includegraphics[width=\figsize]{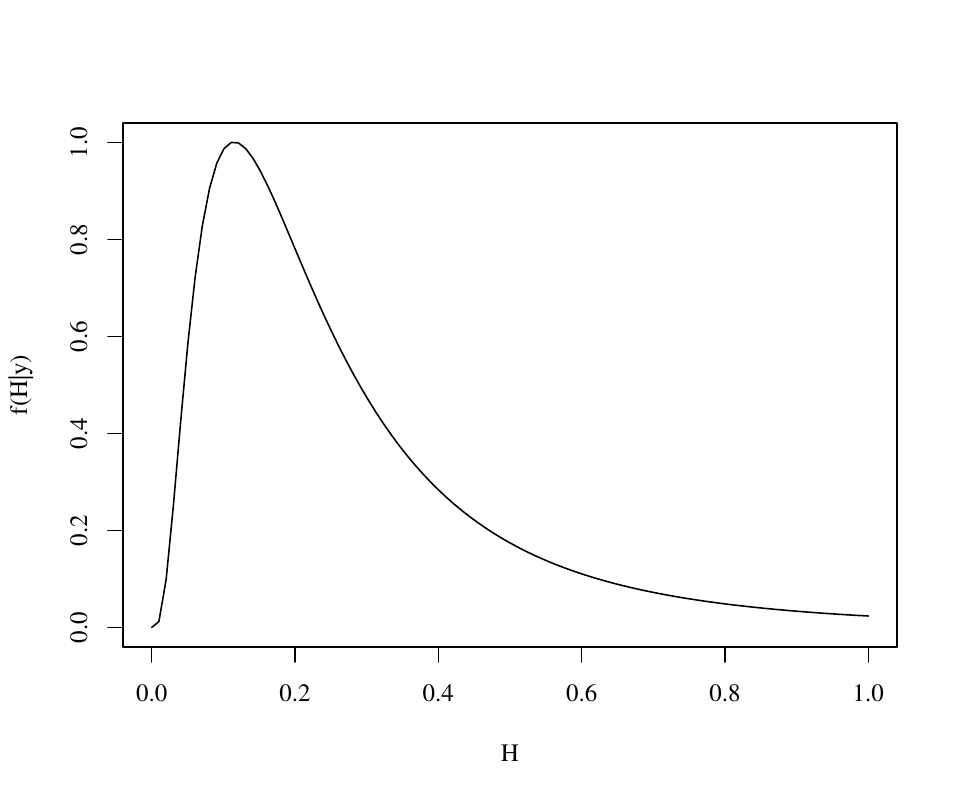}
    \label{fig:plot15}
\end{figure}

\section{Case study: Gastroenteritis in Brazil}\label{Sec:casestudy}

In this example we model number of hospital admissions from gastroenteritis in Brazil from 2010 to 2022 \citep{datasus_data}. The \pkg{kDGLM} package provides the \code{gatroBR} dataset with the pertinent data for our models, which includes:

\begin{itemize}
\item UF: The abbreviated state name.
\item Date: The date of the observation. Note that the day is only a placeholder, as we are dealing with monthly reports.
\item Admissions: The number hospital admissions that were reported in that combination of state and date.
\item Population: The estimated population in that combination of state and date (to be used as a offset). 
\end{itemize}

Supplementary information can be found in the documentation (see \code{help(gastroBR)})).

\subsection{Initial model: Total hospital admissions}\label{initial_model}

We will start with a model for the total number of hospital admissions over time in Brazil, i.e., a temporal model that does not considers the effects of each state.

\begin{figure}[H]
    \centering
    \includegraphics[width=\figsize]{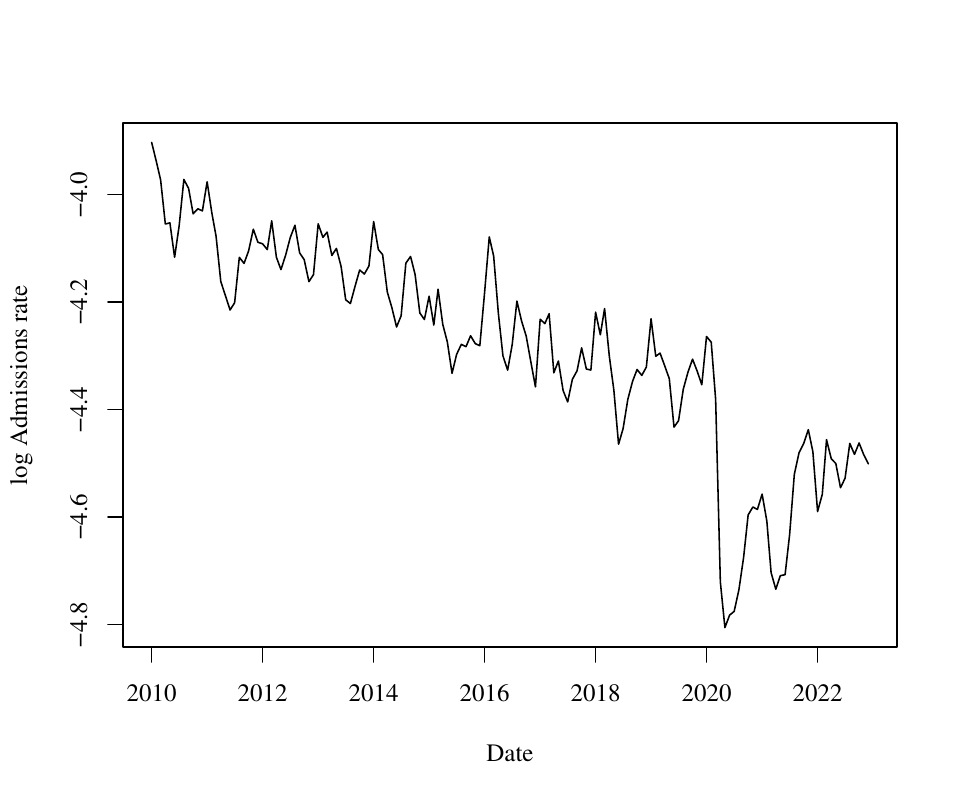}
    \caption{Hospital admissions on Brazil, from 2010 to 2022, on log scale.}
    \label{fig:plot16}
\end{figure}

From figure \ref{fig:plot16} we can see that there is a consistent trend of (log) linear decay of the rate of hospital admissions over time, until April of 2020, when there is an abrupt reduction of hospital admission due to the pandemic of COVID-19 \citep{covid_reduc}, which is then followed by what seems to be a return to the previous level, although more observations would be necessary to be sure. We can also note that the data has a clear seasonal pattern, which have a period of 12 months.

We begin with a very simply model. Let $Y_t$ be the total number of hospital admissions on Brazil at time $t$. Assume that:

$$
\begin{aligned}
Y_t|\eta_t &\sim Poisson(\eta_t)\\
\ln{\eta_t}&=\lambda_t=\theta_{1,t}\\
\theta_{1,t}&=\theta_{1,t-1}+\theta_{2,t-1}+\omega_{1,t}\\
\theta_{2,t}&=\theta_{2,t-1}+\omega_{2,t}.\\
\end{aligned}
$$

First we define the model structure:

\begin{lstlisting}
structure <- polynomial_block(rate = 1, order = 2, D = c(0.95, 0.975),
                              name='Trend')
\end{lstlisting}

Then we define the outcome:

\begin{lstlisting}
outcome <- Poisson(lambda = "rate",
  data = data.year$Admissions,
  offset = data.year$Population)
\end{lstlisting}

Then we fit the model:

\begin{lstlisting}
fitted.model <- fit_model(structure, outcome)
\end{lstlisting}

Finally, we can see how our model performed with the \code{summary} and \code{plot} methods:

\begin{lstlisting}
summary(fitted.model)
\end{lstlisting}

\begin{\verbatimfont} \begin{verbatim}
Fitted DGLM with 1 outcomes.

distributions:
    Series.1: Poisson

Coeficients (smoothed) at time 156:
               Estimate Std. Error   t value Pr(>|t|)
Trend.Level -10.49460    0.00377 -2784.39712   <1e-12 *** 
Trend.Slope  -0.00806    0.00019   -41.95112   <1e-12 *** 
---
Signif. codes:  0 '***' 0.001 '**' 0.01 '*' 0.05 '.' 0.1 ' ' 1

---
One-step-ahead prediction
Log-likelihood        :  -24560.91910
Interval Score        :   52492.80142
Mean Abs. Scaled Error:       1.31369
Relative abs. Error   :       0.17520
Mean Abs. Error       :    1501.65306
Mean Squared Error    :  3.691e+06
 \end{verbatim}\end{\verbatimfont}

\begin{lstlisting}
plot(fitted.model)
\end{lstlisting}
\begin{figure}[H]
    \centering
    \includegraphics[width=\figsize]{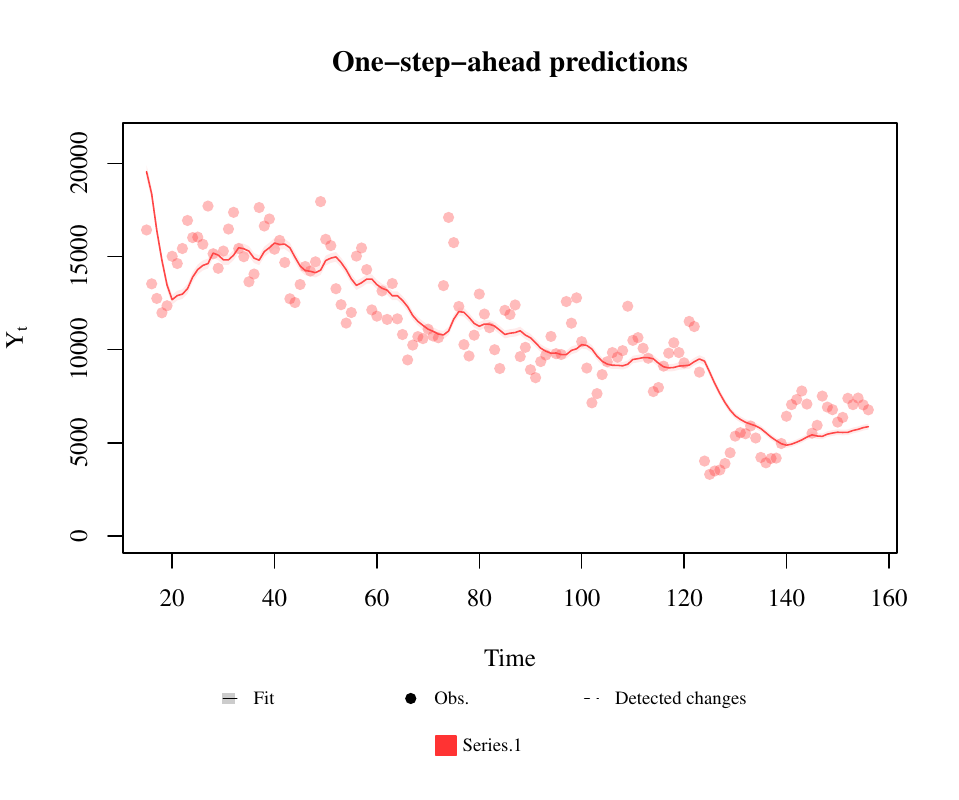}
    \label{fig:plot17}
\end{figure}

Clearly the model described above is too simply to explain the data, in particular, it does not take into account any form of seasonal pattern. Let us proceed then by assuming the following model:

$$
\begin{aligned}
Y_t|\eta_t &\sim Poisson(\eta_t)\\
\ln{\eta_t}&=\lambda_t=\theta_{1,t}+\theta_{3,t}\\
\theta_{1,t}&=\theta_{1,t-1}+\theta_{2,t-1}+\omega_{1,t}\\
\theta_{2,t}&=\theta_{2,t-1}+\omega_{2,t},\\
\begin{bmatrix}\theta_{3,t}\\\theta_{4,t}\end{bmatrix}&=R\begin{bmatrix}\theta_{3,t}\\\theta_{4,t}\end{bmatrix}+\begin{bmatrix}\omega_{3,t}\\\omega_{4,t}\end{bmatrix}
\end{aligned}
$$

Where $R$ is a rotation matrix with angle $\frac{2\pi}{12}$, such that $R^{12}$ is equal to the identity matrix.

To define the structure of that model we can use the \code{harmonic_block} function alongside the \code{polynomial_block} function:

\begin{lstlisting}
structure <- polynomial_block(rate = 1, order = 2, D = c(0.95, 0.975),
                              name='Trend') +
             harmonic_block(rate = 1, period = 12, D = 0.98,
                            name='Season')
\end{lstlisting}

Then we fit the model (using the previously defined outcome):

\begin{lstlisting}
fitted.model <- fit_model(structure, outcome)
summary(fitted.model)
\end{lstlisting}
\begin{\verbatimfont} \begin{verbatim}
Fitted DGLM with 1 outcomes.

distributions:
    Series.1: Poisson

Coeficients (smoothed) at time 156:
               Estimate Std. Error   t value Pr(>|t|)
Trend.Level -10.50216    0.00394 -2663.44603   <1e-12 *** 
Trend.Slope  -0.00788    0.00022   -36.22919   <1e-12 *** 
Season.Main    0.13913    0.00240    57.87650   <1e-12 *** 
Season.Aux     0.02138    0.00235     9.09009   <1e-12 *** 
---
Signif. codes:  0 '***' 0.001 '**' 0.01 '*' 0.05 '.' 0.1 ' ' 1

---
One-step-ahead prediction
Log-likelihood        :  -17311.70077
Interval Score        :   41412.16312
Mean Abs. Scaled Error:       0.64692
Relative abs. Error   :       0.15308
Mean Abs. Error       :    1319.68499
Mean Squared Error    :  2.985e+06
 \end{verbatim}\end{\verbatimfont}
\begin{lstlisting}
plot(fitted.model)
\end{lstlisting}
\begin{figure}[H]
    \centering
    \includegraphics[width=\figsize]{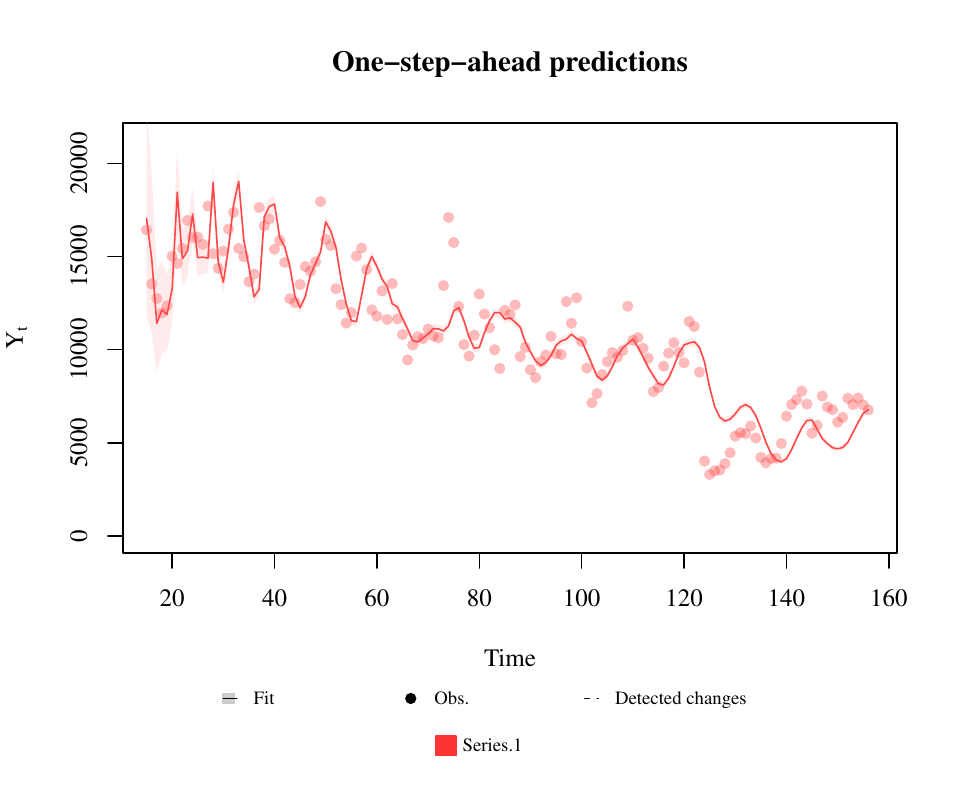}
    \label{fig:plot18}
\end{figure}

Notice that this change significantly improves all metrics provided by the model summary, which indicates that we are going in the right direction. We leave as an exercise for the reader to test different orders for the harmonic block.

The previous model could capture the mean behavior of the series reasonably well, still, two deficiencies of that model standout: First, the overconfidence in the predictions, evidenced by the particularly thin credibility interval; And second, the difficulty the model had to adapt to the pandemic period.

The first problem comes from the fact that we are using a Poisson model, which implies that $Var[Y_t|\eta_t]=\mathbb{E}[Y_t|\eta_t]$, which means that $Var[Y_t]=\mathbb{E}[Var[Y_t|\eta_t]]+Var[\mathbb{E}[Y_t|\eta_t]]=\mathbb{E}[\eta_t]+Var[\eta_t]$. For latter observations we expect that $Var[\eta_t]$ to be relatively small, as such, the variance of $Y_t$ should be very close to its mean after a reasonable amount of observations. In this scenario, the coefficient of variation, defined as $\frac{\sqrt{Var[Y_t]}}{\mathbb{E}[Y_t]}$ goes to $0$ as $\mathbb{E}[Y_t]$ grows, in particular, for data in the scale we are working with in this particular problem, we would expect a very low coefficient of variation if the Poisson model were adequate, but that is not what we observe. This phenomena is called \textit{overdipersion} and is a well known problem in literature \citep[for example]{sari2021handling}. To solve it, we can include a block representing a white noise that is added to the linear predictor at each time, but does not affect previous or future observation, so as to capture the overdipersion. In this case, we will assume the following model:

$$
\begin{aligned}
Y_t|\eta_t &\sim Poisson(\eta_t)\\
\ln{\eta_t}&=\lambda_t=\theta_{1,t}+\theta_{3,t}+\epsilon_t\\
\theta_{1,t}&=\theta_{1,t-1}+\theta_{2,t-1}+\omega_{1,t}\\
\theta_{2,t}&=\theta_{2,t-1}+\omega_{2,t},\\
\begin{bmatrix}\theta_{3,t}\\\theta_{4,t}\end{bmatrix}&=R\begin{bmatrix}\theta_{3,t}\\\theta_{4,t}\end{bmatrix}+\begin{bmatrix}\omega_{3,t}\\\omega_{4,t}\end{bmatrix}\\
\epsilon_t & \sim \mathcal{N}(0,\sigma_t^2)
\end{aligned}
$$

This structure can be defined using the \code{noise_block} function, alongside the previously used functions:

\begin{lstlisting}
structure <- structure + noise_block(rate = 1,name='Noise')
\end{lstlisting}

For the second problem, that of slow adaptation after the start of the pandemic. The ideal approach would be to make an intervention, increasing the uncertainty about the latent states at the beginning of the pandemic period and allowing our model to quickly adapt to the new scenario \citep[see][chapter 11]{WestHarr-DLM}. We recommend this approach when we already expect a change of behavior in a certain time, even before looking at the data (which is exactly the case). Still, for didactic purposes, we will first present how the automated monitoring can also be used to solve this same problem. In general, we recommend the automated monitoring approach when we do \textbf{not} known if or \textbf{when} a change of behavior happened before looking at the data, i.e., we do not known of any particular event that we expect to impact our outcome.

Following what was presented in the subsection \ref{intervention}, we can use the following code to fit our model:

\begin{lstlisting}
structure <- polynomial_block(
  rate = 1, order = 2, D = c(0.95, 0.975),
  name='Trend',monitoring = c(TRUE, TRUE)) +
  harmonic_block(rate = 1, period = 12, D = 0.98,
                 name='Season') +
  noise_block(rate = 1,name='Noise')
# To activate the automated monitoring it is enough to set the p.monit argument to a valid value
fitted.model <- fit_model(structure, outcome, p.monit = 0.05)
\end{lstlisting}

\begin{lstlisting}
structure <- polynomial_block(
  rate = 1, order = 2, D = c(0.95, 0.975), name='Trend') +
  harmonic_block(rate = 1, period = 12, D = 0.98, name='Season') +
  noise_block(rate = 1,R1=0.015,name='Noise')
fitted.model <- fit_model(structure, outcome)
\end{lstlisting}

Notice that we set the \code{monitoring} of the \code{polynomial_block} to \code{c(TRUE,TRUE)}. By default, the \code{polynomial_block} function only activates the monitoring of its first component (the level), but, by the visual analysis made at the beginning, it is clear that the pandemic affected both the level and the slope of the average number of hospital admissions, as such, we would like to monitor both parameters.

\begin{lstlisting}
summary(fitted.model)
\end{lstlisting}

\begin{\verbatimfont} \begin{verbatim}
Fitted DGLM with 1 outcomes.

distributions:
    Series.1: Poisson

Coeficients (smoothed) at time 156:
            Estimate Std. Error   t value Pr(>|t|)
Trend.Level -10.29317    0.06745 -152.60175   <1e-12 *** 
Trend.Slope   0.01513    0.00704    2.15079    0.031 * 
Season.Main   0.10590    0.03732    2.83789    0.005 ** 
Season.Aux   -0.01038    0.03770   -0.27540    0.783   
Noise.Var     0.00000    0.15239    0.00000    1.000   
---
Signif. codes:  0 '***' 0.001 '**' 0.01 '*' 0.05 '.' 0.1 ' ' 1

---
One-step-ahead prediction
Log-likelihood        :   -1260.11029
Interval Score        :   12658.79433
Mean Abs. Scaled Error:       0.63103
Relative abs. Error   :       0.12972
Mean Abs. Error       :    1287.26529
Mean Squared Error    :  3.180e+06
---
 \end{verbatim}\end{\verbatimfont}

\begin{figure}[H]
    \centering
    \includegraphics[width=\figsize]{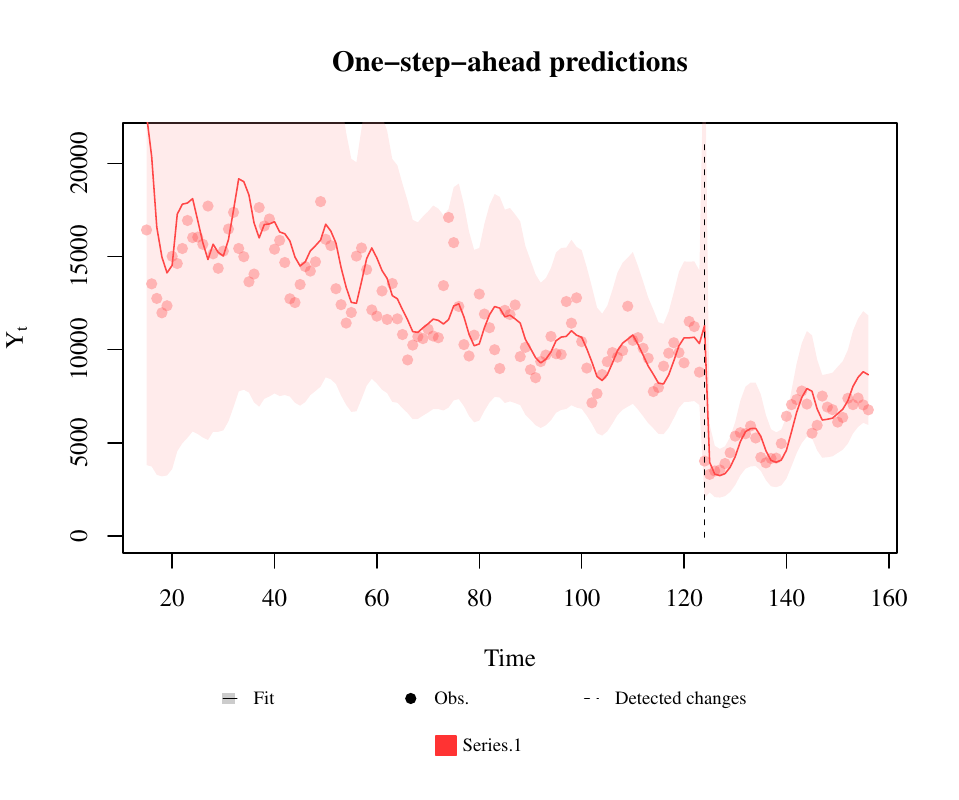}
    \caption{}
    \label{fig:plot19}
\end{figure}

The summary presented above shows a massive improvement in the comparison metrics with the new changes introduced. Moreover, in figure \ref{fig:plot19}, we can see that the automated monitoring detected the exact moment where the series $Y_t$ changed behavior, which allowed the model to immediately adapt to the pandemic period.

\begin{lstlisting}
plot(fitted.model)
\end{lstlisting}

One aspect of the model that may bother the reader is the exceedingly high uncertainty at the first observations. This behavior is duo to our approach to the estimation of the variance of the white noise introduced by the \code{noise_block} function (see \citealp{ArtigoMultivar} and the associated documentation for details), which can be a bit too sensitive to bad prior specification at the initial steps. As such, we highly recommend the user to do a sensibility analysis to choose the initial variance of the white noise:

\begin{lstlisting}
# Now we will make an intervention instead of using the automated monitoring.
structure <- structure |> intervention(time = 124, var.index = c(1:2), D = 0.005)

search.model <- search_model(
  structure, outcome,
  lag = -1, # Using the model likelihood (f(y|M)) as the comparison metric.
  search.grid = list(H = seq.int(0, 0.1, l = 101)))
fitted.model <- search.model$model
\end{lstlisting}

Notice that, this time around, we chose to  make an intervention at the beginning of the pandemic, instead of an automated approach. As mentioned before, this approach is preferable in this scenario, since we were aware that the pandemic would affect our outcome before even looking at the data.

\begin{lstlisting}
summary(fitted.model)
\end{lstlisting}

\begin{\verbatimfont} \begin{verbatim}
Fitted DGLM with 1 outcomes.

distributions:
    Series.1: Poisson

Coeficients (smoothed) at time 156:
               Estimate Std. Error   t value Pr(>|t|)
Trend.Level -10.28202    0.04914 -209.24953   <1e-12 *** 
Trend.Slope   0.01624    0.00490    3.31793 9.07e-04 *** 
Season.Main    0.10444    0.02339    4.46500 8.01e-06 *** 
Season.Aux     0.00504    0.02350    0.21436    0.830   
Noise.Var        0.00000    0.11451    0.00000    1.000   
---
Signif. codes:  0 '***' 0.001 '**' 0.01 '*' 0.05 '.' 0.1 ' ' 1

---
One-step-ahead prediction
Log-likelihood        :   -1221.04357
Interval Score        :    7317.21986
Mean Abs. Scaled Error:       0.54414
Relative abs. Error   :       0.11413
Mean Abs. Error       :    1110.01383
Mean Squared Error    :  2.351e+06
 \end{verbatim}\end{\verbatimfont}

Again, the new changes improve the comparison metrics even further, leading to the conclusion that our last model is the best among those presented until now. We highly encourage the reader to run this example and experiment with some of the options the \pkg{kDGLM} package offers, but that were not explored, such as changing the discount factors used in each block, the order of the blocks, adding/removing structural components, etc..

\begin{lstlisting}
plot(fitted.model)
\end{lstlisting}

\begin{figure}[H]
    \centering
    \includegraphics[width=\figsize]{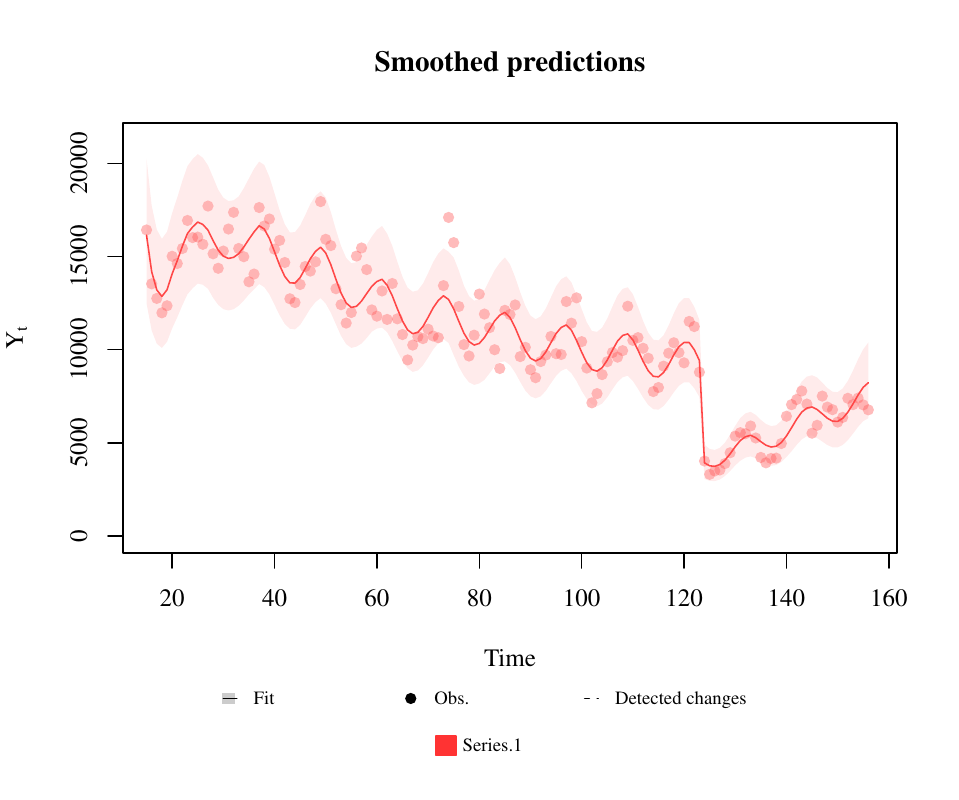}
    \label{fig:plot20}
\end{figure}

As a last side note, the user may not like the approach of choosing a specific value for the initial variance of the white noise introduced by the \code{noise_block}. Indeed, one may wish to define a prior distribution for this parameter and estimate it along with the others. While we will not detail this approach for the sake of brevity (since it is not directly supported), we would like to point out that we do offer tools to facilitate this procedure:

\begin{lstlisting}
search.result <- search.model$search.data[order(search.model$search.data$H), ]

H.vals <- search.result$H
log.prior <- dgamma(H.vals, 1, 1, log = TRUE)
log.like <- search.result$log.like
l.fx <- log.prior + log.like
pre.fx <- exp(l.fx - max(l.fx))
fx <- pre.fx / sum(pre.fx * (H.vals[2] - H.vals[1]))
plot(H.vals, fx,
  type = "l", xlab = "H", ylab = "Density",
  main = "Posterior density for the unknown hyperparameter H")
\end{lstlisting}

\begin{figure}[H]
    \centering
    \includegraphics[width=\figsize]{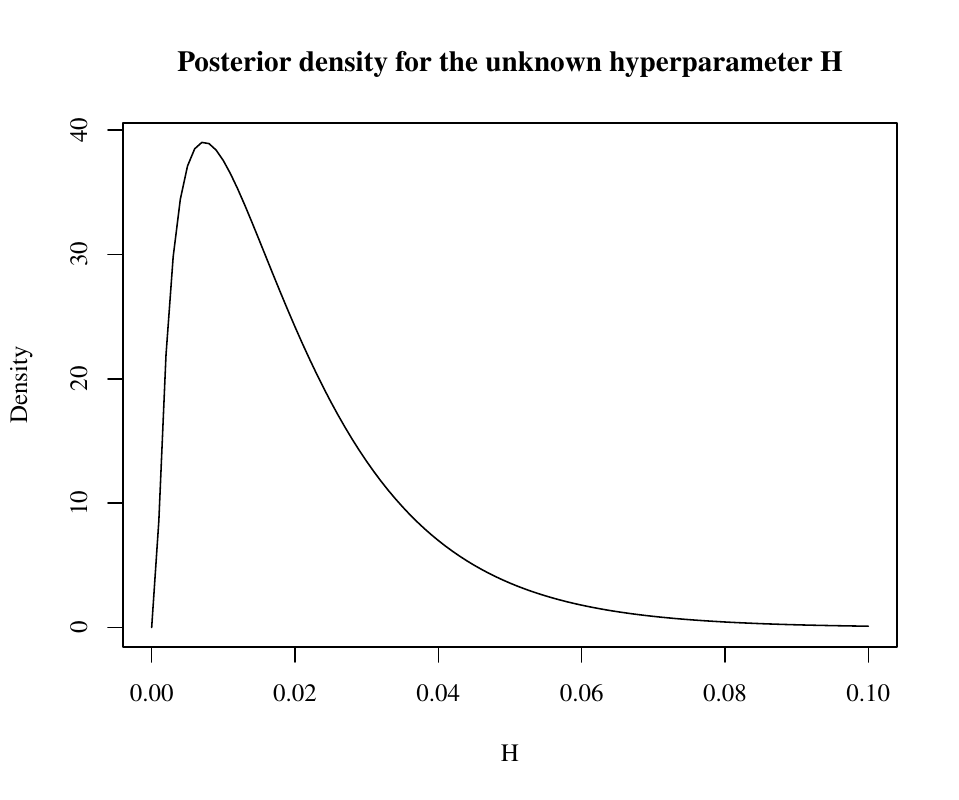}
    \label{fig:plot21}
\end{figure}

\subsection{Advanced model: Hospital admissions by state}

For this model, we need the geographic information about Brazil, as such, we will use some auxiliary packages, namely \pkg{geobr}, \pkg{tidyverse}, \pkg{sf} and \pkg{spdep}, although the \pkg{kDGLM} package does not depend on them:

\begin{lstlisting}
require(geobr); require(tidyverse); require(sf); require(spdep)

br.base <- read_state(year = 2019, showProgress = FALSE)
plot.data <- br.base |>
  left_join(gastroBR |>
      filter(format(Date, "%Y") == "2019") |>
      select(UF, Population, Admissions) |>
      group_by(UF) |>
      summarize(
        Population = max(Population),
        Admissions = sum(Admissions)) |>
      rename(abbrev_state = UF),
    by = "abbrev_state")

ggplot() +
  geom_sf(data = plot.data, aes(fill = log10(Admissions / Population))) +
  scale_fill_continuous("") +
  theme_void()
\end{lstlisting}

\begin{figure}[H]
    \centering
    \includegraphics[width=\figsize]{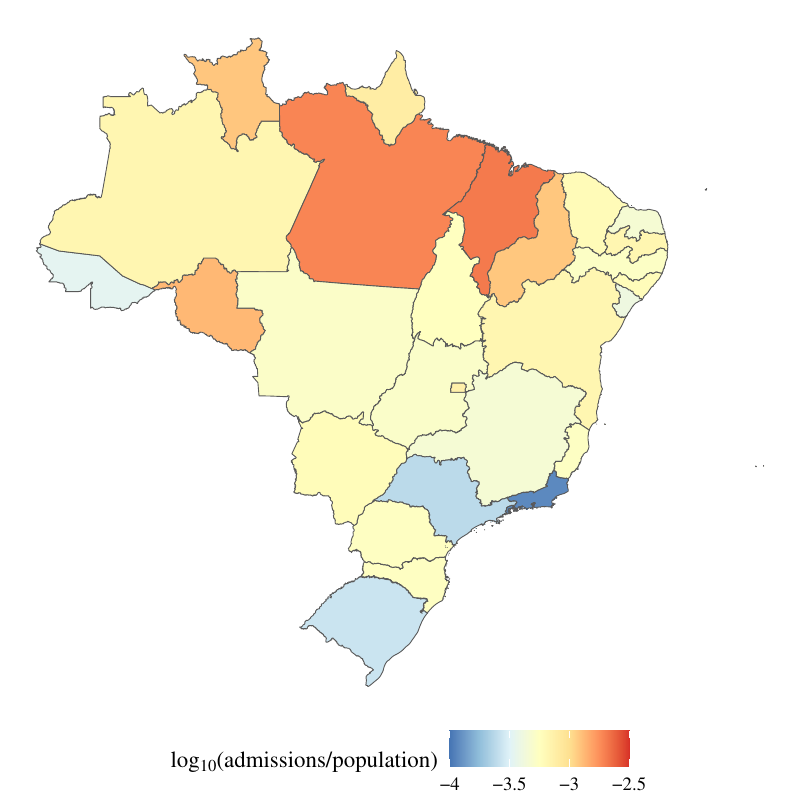}
    \label{fig:plot22}
    \caption{The $\log_{10}$ rate of hospital admission on Brazil by state. Here we consider the total number of hospital admission, i.e., we agregate all years.}
\end{figure}

Now we proceed to fitting the model itself. Let $Y_{it}$ be the number of hospital admissions by gastroenteritis at time $t$ on region $i$, we will assume the following model:

$$
\begin{aligned}
Y_{it}|\eta_{it} &\sim Poisson(\eta_{it})\\
\ln\{\eta_{it}\}&= \lambda_{it}=\theta_{1,t}+u_{i,t}+S_{i,t}+\epsilon_{i,t},\\
\theta_{1,t}&= \theta_{1,t-1}+\theta_{2,t-1}+\omega_{1,t},\\
\theta_{2,t}&= \theta_{2,t-1}+\omega_{2,t},\\
\begin{bmatrix}u_{i,t}\\ v_{i,t}\end{bmatrix} &= R \begin{bmatrix}u_{i,t-1}\\ v_{i,t-1}\end{bmatrix} + \begin{bmatrix} \omega^{u}_{i,t}\\ \omega^{u}_{i,t}\end{bmatrix},\\
\epsilon_t & \sim \mathcal{N}(0,\sigma_t^2),\\
S_{1,1},...,S_{r,1} & \sim CAR(100),
\end{aligned}
$$
where $r=27$ is the number of areas within our dataset.

Notice that we are assuming a very similar model to that which was used in subsection \ref{initial_model}, but here we have a common effect (or a global effect) $\theta_{1,t}$ that equally affects all regions, and a local effect $S_{i,t}$ that only affects region $i$ and evolves smoothly though time. Here we chose a vague CAR prior \citep{AlexCar,banerjee2014hierarchical} for $S_{i,t}$.


The proposed model can be fitted using the following code.

\begin{lstlisting}
adj.matrix <- ref.br |> poly2nb() |> nb2mat(style = "B")

CAR.structure <- polynomial_block(rate = 1, D = 0.8, name = "CAR") |>
  block_mult(27) |> block_rename(levels(gastroBR$UF)) |>
  CAR_prior(scale = 9, rho = 1, adj.matrix = adj.matrix)

shared.structure <- polynomial_block(
  RO = 1, AC = 1, AM = 1, RR = 1, PA = 1, AP = 1, TO = 1, MA = 1, PI = 1,
  CE = 1, RN = 1, PB = 1, PE = 1, AL = 1, SE = 1, BA = 1, MG = 1, ES = 1,
  RJ = 1, SP = 1, PR = 1, SC = 1, RS = 1, MS = 1, MT = 1, GO = 1, DF = 1,
  order = 2, D = c(0.95, 0.975), name = "Common") |>
  intervention(time = 124, var.index = c(1:2), D = 0.005)

base.structure <- (harmonic_block(rate = 1, period = 12, D = 0.98,
                   name = "Season") +
  noise_block(rate = 1, R1 = 0.007, name = "Noise")) |>
  block_mult(27) |> block_rename(levels(gastroBR$UF))

inputs <- list(shared.structure, CAR.structure, base.structure)
for (uf in levels(gastroBR$UF)) {
  reg.data <- gastroBR %>% filter(UF == uf)
  inputs[[uf]] <- Poisson(lambda = uf, data = reg.data$Admissions,
                          offset = reg.data$Population)}
fitted.model <- do.call(fit_model, inputs)
\end{lstlisting}

\begin{figure}[H]
    \centering
    \includegraphics{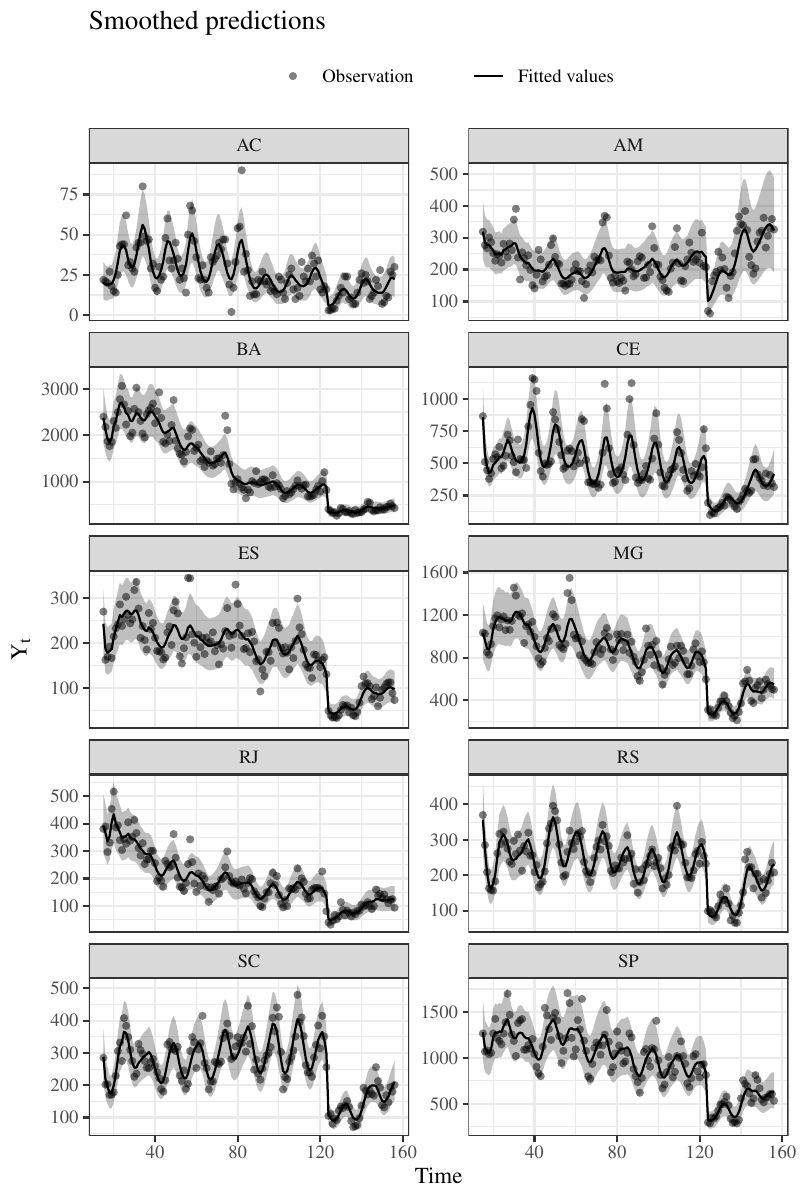}
    \caption{The time series of hospital admissions by gastroenteritis of some Brazilian states from 2010 to 2022. Notice that the proposed model can capture the general behavior of all series, while simultaneously capturing the dependence between regions through the shared component $\theta_{1,t}$ and the local effects $S_i$.}
    \label{fig:plot23}
\end{figure}

\begin{figure}[H]
    \centering
    \includegraphics[trim={0 5cm 0 5cm}]{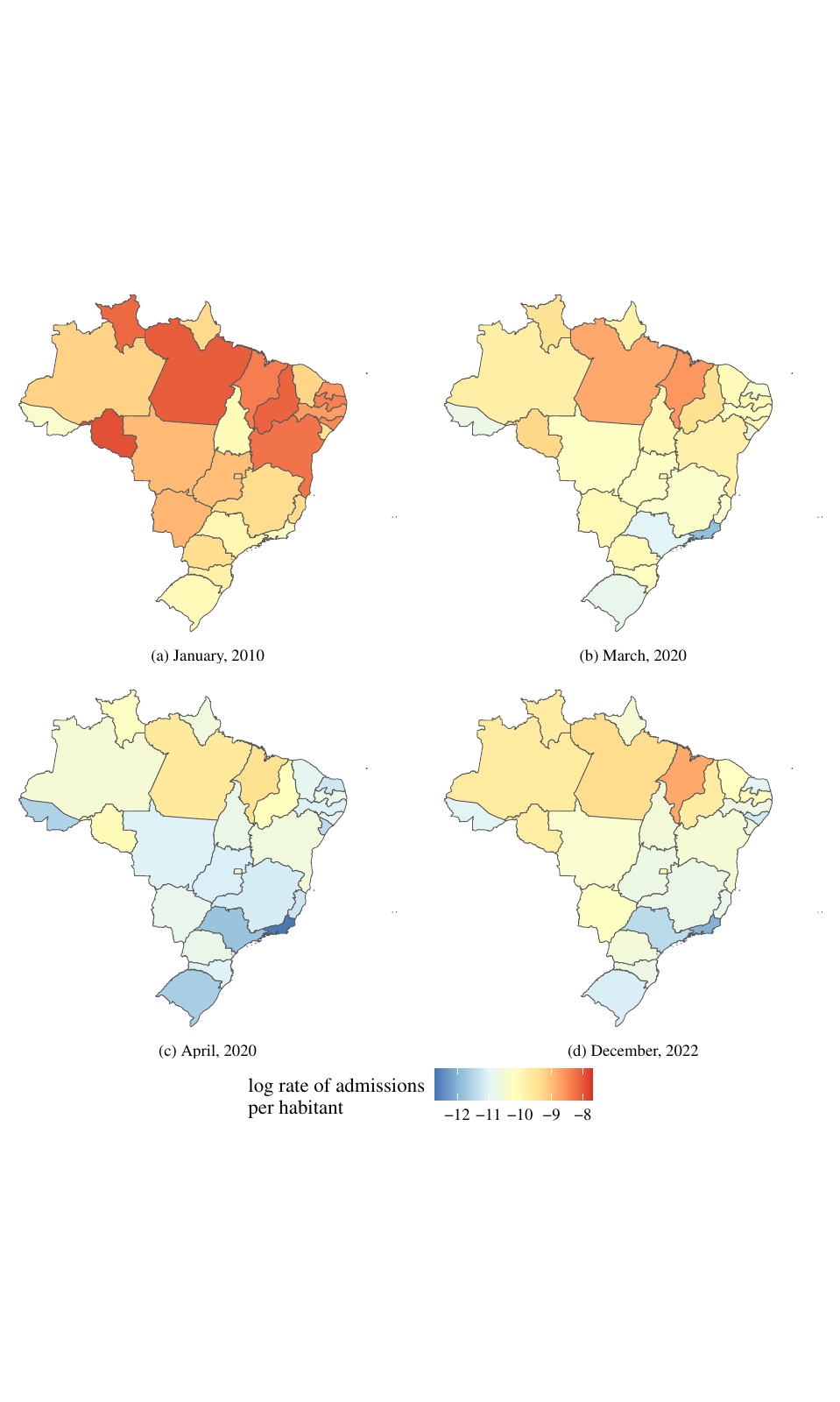}
    \caption{The $\log_{10}$ hospital admissions rate by gastroenteritis in Brazilian states at 4 key moments: (a) January of 2010, were our data begins; (b) March of 2020, the month were the first case of COVID-19 was registered in Brazil and before public response; (c) April of 2020, the first month of the pandemic period; and (d) December of 2022, the end of the period of study and roughly 2 years after the beginning of the pandemic. Notice that from (a) to (b) 10 years had passed and we see that a steady and smoothly yearly reductions of hospital admissions led to a significantly reduction of the rate of hospital. In contrast, from (b) to (c), only 1 month had passed, but we see a reduction that, proportionally, is event greater than from (a) to (b). Lastly, from (c) to (d), after roughly 2 years, the rate of hospital admissions seems to be going back to what was seen in (c).}
    \label{fig:plot24}
\end{figure}

Since there are $27$ regions, with $156$ observations each, its not reasonably to show how our model performed for every combination of date and location. We will limit ourselves to show some regions at all times (figure \ref{fig:plot23}) and all region at some times (figure \ref{fig:plot24}). The reader may use the code provided in this document or in the vignette to fit this model and do a thoroughly examination of the results. Moreover, here we focus only in the usage of the \pkg{kDGLM} package and not in the epidemiological aspect of the results.

\begin{lstlisting}
# Code for figure 4
plot(fitted.model,lag = -1, plot.pkg = "ggplot2",
     outcomes=c('MG','SP','ES','RJ','CE','BA','RS','SC','AM','AC')) +
scale_color_manual('',values=rep('black',10)) +
scale_fill_manual('',values=rep('black',10)) +
facet_wrap(~Serie, ncol = 2, scale = "free_y") +
coord_cartesian(ylim = c(NA, NA)) +
guides(color = "none", fill = "none") + theme(legend.position = "top")
# Code for figure 5
CAR.index <- which(grepl("CAR", fitted.model$var.labels))
plot.data <- data.frame()
labels=list('2010-01-01'='(a) January, 2010', '2020-03-01'='(b) March, 2020',
            '2020-04-01'='(c) April, 2020', '2022-12-01'='(d) December, 2022')
for (date in names(labels)) {
  i <- which(gastroBR$Date == date)[1]
  plot.data <- plot.data |> rbind(cbind(Date = labels[[date]], br.base,
  Effect = fitted.model$mts[1, i] + fitted.model$mts[CAR.index, i]))}

(ggplot() + geom_sf(data = plot.data, aes(fill = Effect)) + 
  facet_wrap(~Date, strip.position="bottom") +
  scale_fill_distiller("log rate of admissions\nper habitant",
    limits = c(-12.7, -7.7), palette = "RdYlBu", labels = ~ round(., 2)) +
  theme_void() + theme(legend.position = "bottom"))
\end{lstlisting}

Finally, about the computational cost, the initial model (that for the total number of hospital admissions over time) took about $0.11s$ to fit and the advanced model took $4.24s$, which is within the expected range, since the final model has $27$ outcomes and $110$ latent states that, when we consider that they all had temporal dynamic, yields $17.160$ parameters, from which the joint distribution is obtained. 

\section{Conclusion and Future Directions}\label{Sec:Conclusion}

In conclusion, the \pkg{kDGLM} package provides a robust and efficient framework for Bayesian analysis of Generalized Dynamic Linear Models, offering powerful tools for the exploration of time series data. Its seamless integration with established techniques, coupled with low computational costs, makes it an appealing choice for researchers and practitioners dealing with extensive temporal datasets, specially for problems of real-time monitoring and online inference. 

Looking ahead, the development of \pkg{kDGLM} is an ongoing process. The package will continue to evolve with the incorporation of additional probability distributions and model structures to enhance its versatility. Future updates will focus on expanding the range of supported models and further refining the computational efficiency. User feedback and contributions will play a crucial role in shaping the trajectory of the \pkg{kDGLM} package, ensuring its adaptability to diverse research needs.

As the field of Bayesian time series analysis progresses, \pkg{kDGLM} aims to remain at the forefront, providing an accessible and powerful toolset for researchers across various domains, while incorporating future new developments in the Bayesian inference. The package's user-friendly interface, coupled with its computational speed, positions it as a valuable asset for both seasoned Bayesian statisticians and those new to the field.

In conclusion, \pkg{kDGLM} stands as a testament to the intersection of theoretical advancements and practical applicability in Bayesian time series analysis, promising continuous innovation and adaptation to meet the evolving demands of the research community.

\section{Acknowledgments}

The authors thank Raíra Marotta (UFRJ?), for the initial implementation of the methodology proposed in \cite{ArtigokParametrico}, which was the starting point of the \pkg{kDGLM} package.

\pagebreak

\bibliographystyle{apalike}
\bibliography{references}

\end{document}